\documentclass{article}\usepackage[]{graphicx}\usepackage[]{xcolor}
\makeatletter
\def\maxwidth{ %
  \ifdim\Gin@nat@width>\linewidth
    \linewidth
  \else
    \Gin@nat@width
  \fi
}
\makeatother

\definecolor{fgcolor}{rgb}{0.345, 0.345, 0.345}

\usepackage{framed}
\makeatletter
\newenvironment{kframe}{%
 \def\at@end@of@kframe{}%
 \ifinner\ifhmode%
  \def\at@end@of@kframe{\end{minipage}}%
  \begin{minipage}{\columnwidth}%
 \fi\fi%
 \def\FrameCommand##1{\hskip\@totalleftmargin \hskip-\fboxsep
 \colorbox{shadecolor}{##1}\hskip-\fboxsep
     \hskip-\linewidth \hskip-\@totalleftmargin \hskip\columnwidth}%
 \MakeFramed {\advance\hsize-\width
   \@totalleftmargin\z@ \linewidth\hsize
   \@setminipage}}%
 {\par\unskip\endMakeFramed%
 \at@end@of@kframe}
\makeatother

\definecolor{shadecolor}{rgb}{.97, .97, .97}
\definecolor{messagecolor}{rgb}{0, 0, 0}
\definecolor{warningcolor}{rgb}{1, 0, 1}
\definecolor{errorcolor}{rgb}{1, 0, 0}
\newenvironment{knitrout}{}{} 

\usepackage{alltt}

\usepackage{arxiv}
\usepackage[utf8]{inputenc} 
\usepackage[T1]{fontenc}    
\usepackage{hyperref}       
\usepackage{booktabs}       
\usepackage{amsfonts}       
\usepackage{multicol}

\usepackage{array}
\usepackage{float}
\usepackage{framed}
\usepackage{amsmath, amsthm,amssymb}
\usepackage{parskip}
\usepackage{listings}
\usepackage{enumitem}
\usepackage{setspace}
\usepackage[english]{babel}
\usepackage{cleveref}

\usepackage{graphicx}
\usepackage{pdflscape}
\usepackage{caption, subcaption}

\newcommand{\proglang}[1]{\textsf{#1}}
\def\bigr{{\mathbb{R}}}
\newcommand{\pkg}[1]{\textbf{\emph{#1}}}
\newcommand{\code}[1]{\texttt{#1}}
\newcommand{\fct}[1]{\textit{#1()}}

\def\cov{{\mathrm{COV}}}
\def\Pr{{\mathbb{P}}}
\def\T{{\footnotesize {^{_{\sf T}}}}}
\def\E{\mathbb{E}} 
\newcommand{\mathleft}{\@fleqntrue\@mathmargin0pt}

\newtheorem*{theorem*}{Theorem}

\title{Profile likelihoods for parameters in trans-Gaussian geostatistical models}
\date{}

\author{
  Ruoyong Xu\\
  Department of Statistical Sciences\\
  University of Toronto\\
  700 University Ave., Toronto, ON M5G 1Z5, Canada\\
  \texttt{ruoyong.xu@mail.utoronto.ca} \\
   \And
  Patrick Brown\\
  Department of Statistical Sciences\\
  University of Toronto\\
  700 University Ave., Toronto, ON M5G 1Z5, Canada\\
  \texttt{patrick.brown@utoronto.ca} \\
}
\IfFileExists{upquote.sty}{\usepackage{upquote}}{}
\begin{document}
\maketitle
\setstretch{1.1}

\begin{abstract}
Profile likelihoods are rarely used in geostatistical models due to the computational burden imposed by repeated decompositions of large variance matrices. 
Accounting for uncertainty in covariance parameters can be highly consequential in geostatistical models as some covariance parameters are poorly identified, 
the problem is severe enough that the differentiability parameter of the Matern correlation function is typically treated as fixed.  
The problem is compounded with anisotropic spatial models as there are two additional parameters to consider.  
In this paper, we make the following contributions:
1, A methodology is created for profile likelihoods for Gaussian spatial models with Mat\'ern family of correlation functions, including anisotropic models. This methodology adopts a novel reparametrization for generation of representative points, and uses GPUs for parallel
profile likelihoods computation in software implementation.\\
2, We show the profile likelihood of the Mat\'ern shape parameter is often quite flat but still identifiable, it can usually rule out very small values. \\
3, Simulation studies and applications on real data examples show that profile-based confidence intervals of covariance parameters and regression parameters have superior coverage to the traditional standard Wald type confidence intervals. \\


\end{abstract}


\section{Introduction}\label{section1}
Anisotropic spatial models with the Mat\'ern family of correlation functions are important but difficult to implement because they involve many model parameters. In addition to the five correlation parameters range ($\phi_X$), shape ($\kappa$), nugget ($\nu^2$), anisotropy ratio ($\phi_R$), anisotropy angle ($\phi_A$), there are the variance parameter ($\sigma$), the Box-Cox transformation parameter ($\lambda$) and $p$ regression parameters $\beta=(\beta_1,\dots,\beta_p$) to estimate. Profile likelihood can be used to reduce the dimensionality of the full likelihood, and it is generally advised to plot the profile likelihood functions for intractable parameters. In spatial models, the likelihoods of the correlation parameters can be far from Normal (e.g., the profile likelihood surface of $\kappa$ can often be very flat), and the true values of the shape, nugget or variance parameter can sometimes be close to a boundary (e.g., true value of $\kappa$ can be infinity). In these situations the curvature (related to the Fisher information matrix) of the likelihood at the maximum likelihood estimates (MLEs) is not a useful approximation of the rest of the likelihood surface, and standard Wald confidence intervals (CIs) are inaccurate.

Most spatial software packages do not use profile likelihoods due to the computational burden imposed by repeated likelihood evaluations. Let $\omega=(\phi_X,\kappa,\nu^2, \phi_{Y}, \phi_{A})$ be the vector of correlation parameters, let $\psi = (\omega, \sigma, \lambda)$ and $\beta_i, i \in \{1,p\}$ denote an element of $\beta$. 
The  geostatistics \proglang{R} package \pkg{geostatsp} 
fixes the parameters $\psi$ at their MLEs, denoted by $\hat{\psi}$, and uses the pseudo-likelihood $\ell(\beta_i; \hat{\psi})$ to evaluate each $\beta_i$. For confidence intervals (CIs) of the covariance parameters, second derivative of the profile likelihood $\ell_p(\psi, \hat{\beta}_{\psi}; y)$ is taken with respect to $\psi$ numerically to obtain the information matrix evaluated at $\hat{\psi}$. Likelihood-based CIs would be advantageous in geostatistical context as the `boundary problem' and `not regular likelihood' often occurs. Obtaining likelihood-based CIs is computed by maximizing over $\psi$,
that is computing the profile likelihood $\ell_p(\beta_i) =\ell(\beta_i, \hat{\psi}_{\beta_i})$ and then deriving CIs from the profile likelihood curves for each $\beta_i$. There is a function
in \pkg{geostatsp} that is able to compute likelihood-based CIs for some of the model parameters ($\lambda$ and $\omega$ only). 
For a fixed value of $\phi_X$, for example, \pkg{geostatsp} 
numerically optimize the joint likelihood of the remaining parameters $(\kappa, \nu^2, \phi_{R}, \phi_{A})$. The process is repeated for a large number of values of $\phi_X$, then it uses the interpolated profile likelihood obtained by interpolating 
to compute likelihood-based CIs for $\phi_X$. This procedure requires running an optimization
function a great many of times so can be very time-consuming, and can fail if the spatial data has a large number of observations. Obtaining plug-in Wald CIs is the relatively `easy way', as quantities (MLEs and information matrix) needed to be calculated are all available from the optimizer 
used to find the MLEs of the correlation parameters. However, $\ell(\beta_i; \hat{\psi})$ does not allow for uncertainty in the estimated correlation parameters to be reflected in the CIs for $\beta_i$'s, hence \pkg{geostatsp}'s CIs for regression parameters would be theoretically narrower than the likelihood-based CIs of $\beta_i$'s. Accounting for uncertainty in correlation parameters is highly consequential in geostatistical models as some correlation parameters are poorly identified, the problem is severe enough that the differentiability parameter $\kappa$ of the Mat\'ern correlation is typically treated as fixed at some plausible value. 

In this paper we present a method that is able to compute the profile likelihoods and likelihood-based CIs for all the parameters in Gaussian linear geostatistical models (LGMs) (anisotropic or isotropic), with dense covariance matrices (Mat\'ern), without approximating the covariance matrix. Our method largely reduces the computation burden of profile likelihoods by taking advantage of 
the parallel computing power of graphic processing units (GPUs), so that the repetitive model fitting at a large number of different parameter values can be done parallely on GPUs, and expedited by use of GPU's local memory.
The method is implemented and made available in the \proglang{R} package \pkg{gpuLik}. 


The rest of this paper is organized as follows.
In section 2, the necessary background of our method is provided. 
Section 3 describes the main idea and procedure and computational work of our method in detail.
In Section 4 we discuss briefly the considerations in GPU computation, and also explained how the computational work is parallelized on a GPU.
Section 5 presents 2 simulation studies. 
In section 6 we demonstrate the applicability of our method by employing it on small and large real data sets. 
Section 7 concludes with a discussion and future directions.
Description on likelihood inference using restricted maximum likelihood for linear geostatistical models is deferred to the Appendix.

\section{Preliminaries}
\label{section2}
The fit of the linear geostatistical model can often be improved by applying a transformation to the observations $Y_i, i=1,\dots,n$. Skewed random fields that deviate from the Normal distribution can be modeled by means of the Box-Cox transformation (BCT) \cite{box1964analysis}. For positive-valued observations $Y_i$, the Box-Cox transformed response $Y_i'$ has the form
\begin{equation*}
	Y_i'= b(Y_i;\lambda) = 
	\begin{cases}
		(Y_i^\lambda-1)/\lambda: & \lambda \in \bigr \quad \text{and} \quad \lambda \neq  0, \\
		\log (Y_i): & \lambda =0.
	\end{cases}
\end{equation*}
The aim of the BCT is to make data closely resemble the Normal distribution, so that the usual assumptions for linear models hold. The BCT cannot guarantee that any probability distribution can be normalized. However, even in cases when an exact mapping to the Normal distribution is not possible, BCT may still provide a useful approximation \cite{hristopulos2020random}.
Consider a zero-mean Gaussian random field $U=\left[U(s_1), \dots,U(s_n)\right]^\T$, 
the Box-Cox transformed responses $Y'_i, i=1,\dots,n$ obtained at locations $s_i$ with explanatory variables $X(s_i)$ are modeled as 
\begin{gather*}
	Y'_{i}|U(s_i) \overset{ind}\sim \text{N}(\mu(s_{i}),\tau^2 ),\\
	\mu(s_{i})=X(s_i)\beta+U(s_{i}).
\end{gather*}
Given $U(s_i)$, $Y_i'$'s are mutually independently distributed with Normal distribution with observation variance $\tau^2$.  
$\beta=(\beta_1,\dots,\beta_p)^\T$ is the vector of regression parameters. 
The covariance between locations $s+h$ and $s$ takes the form below, where $\rho(\cdot)$ is the correlation function, and $\sigma^2$ is the spatial variance or variability in residual variation.  
\begin{gather*}
	\cov \left[U(s+h), U(s)\right]= \sigma^2 \rho(h;\phi,\kappa)
\end{gather*}
Let $Y'=(Y'_1,\dots,Y'_n)^\T$ be the vector of transformed observations, and $X = \left[X(s_1), \dots, X(s_n)\right]^\T$ be an $n \times p$ design matrix, we rewrite the model in compact format
\begin{gather*} 
	Y' \sim \text{MVN} \left(X\beta, \sigma^2 R(\phi,\kappa)+\tau^2 I \right),\\
	R_{ij} = \rho(s_i-s_j;\phi,\kappa),
\end{gather*}
here $R(\cdot)$ is the correlation matrix. Write $V=R(\phi,\kappa)+\nu^2I$, and $\nu^2={\tau^2}/{\sigma^2}$, then
\begin{equation} \label{eq:1}
	Y' \sim \text{MVN} \left(X\beta, \sigma^2 V\right).
\end{equation}



\subsection{Mat\'ern correlation and Geometric anisotroy}
Stein \cite{stein1999interpolation} strongly recommended using the Mat\'ern class of spatial correlations as this class has a parameter which controls the differentiability of the process. 
We use the following parametrization of the Mat\'ern, which includes a rotation parameter to model directional effects.  

\label{eq:matern}
\begin{align}
	\rho(x;\phi,\kappa)= & \frac{2^{\kappa-1}}{\Gamma(\kappa)} 
	\left[\sqrt{8\kappa} d(x) \right]^\kappa  
	K_\kappa\left[ \sqrt{8\kappa}  d(x) \right]\\
	d(x)  = & \left|\left|
	\begin{pmatrix}
		1/\phi_X & 0 \\
		0 & 1 / \phi_Y
	\end{pmatrix}
	\cdot
	\begin{pmatrix}
		\cos \phi_A & -\sin \phi_A \\
		\sin \phi_A & \cos \phi_A 
	\end{pmatrix}
	\cdot
	\begin{pmatrix}
		x_1 \\
		x_2 
	\end{pmatrix}
	\right|\right|
\end{align}

Here  $K_\kappa(\cdot)$ denotes the modified Bessel function of the second kind of order $\kappa$, 
$\kappa$ is the shape parameter that determines the differentiability of $U(x)$. For $\kappa=0.5$, the Mat\'ern correlation function coincides with the exponential correlation $\exp(-d)$. When $\kappa \rightarrow \infty$, it tends to the Gaussian correlation $\exp(-2 d^2)$.  The standard isotropic correlation has a (scalar) range parameter $\phi$ which controls how quickly correlation between two locations falls as the distance between them increases.  The anisotropic model has three parameters which determine distance between points.  The rotation parameter $\phi_A$ gives the direction the correlation is strongest, and the two range parameters $\phi_X$ and $\phi_Y$ which control the speed of decline in the strongest and weakest direction.  The isotropic model results when $\phi_X = \phi_Y$, in which case the angle $\phi_A$ has no effect.

\subsection{Inference based on profile likelihoods}\label{inference}

Let $y'=(y'_1,\dots,y'_n)^\T$ be the Box-Cox transformation of the observations $y=(y_1,\dots,y_n)^\T$.  Further, write $\omega = (\phi_X,\kappa,\nu^2, \phi_{Y}, \phi_{A})$ as the vector of covariance parameters. 
The log-likelihood function for the LGM parameters given $y$ is
\begin{align}
	-2\ell(\beta,\sigma^2, \lambda, \omega; y)=&(y'-X\beta)^\T(\sigma^2 V)^{-1}(y'-X\beta) +\log |\sigma^2 V|  \notag \\
	 &-2*(\lambda-1)\sum_{i=1}^{n}\log y_i+n\log(2\pi),\label{eq:2}
\end{align}
the third term on the right-hand side of \eqref{eq:2} is the Jacobian of the Box-Cox transformation.

Given $\omega$ and $\lambda$, The log-likelihood function \eqref{eq:2} is maximized at
\begin{gather}
	\hat{\beta}_{\omega,\lambda}=(X^\T V^{-1}X)^{-1} X^\T V^{-1}y', \quad \text{and} \label{betahat}\\
	\hat{{\sigma}}^2_{\omega,\lambda}= \frac{1}{n}(y'-X\hat{\beta}_{\omega,\lambda})^\T V^{-1}(y'-X\hat{\beta}_{\omega,\lambda}) \label{eq:4}.
\end{gather}

Substituting \eqref{betahat} and \eqref{eq:4} into \eqref{eq:2}, we obtain the profile log-likelihood (PLL) for $\omega$ and $\lambda$
\begin{align} \label{profile}
	-2\ell_p(\omega, \lambda;y)=&n \log\frac{(y'-X\hat{\beta}_{\omega,\lambda})^\T V^{-1}(y'-X\hat{\beta}_{\omega,\lambda})}{n}+\log|V|  \notag \\
	&-2(\lambda-1)\sum_{i=1}^{n}\log y_i + n\log(2\pi) + n.
\end{align}
There is no closed formula for the MLE's of the model parameters.  Numerical optimization of \eqref{profile}  yields the MLE's $\hat\omega$ and $\hat\lambda$, and back substituting them into \eqref{betahat} and \eqref{eq:4} leads to the MLE's $\hat\beta$ and $\hat\sigma$. Inference for the model parameters using restricted maximum profile likelihood is briefly presented in the Appendix.

\subsection{Information-based and likelihood-based confidence intervals}

We use $\hat{\theta}$ to denote the MLE of an arbitrary parameter vector $\theta = (\theta_1 \ldots \theta_q)$.  Assymptotic $100(1 - \alpha)\%$ confidence intervals for a single parameter $\theta_i$ can be constructed in two ways.
\begin{description}[style=unboxed,leftmargin=0cm]
	\item[1. Wald, or information-based confidence interval] 
	\begin{equation}\label{wald}
		\{\theta_i: |\theta_i - \hat{\theta_i}|  \leq z_{1-\alpha/2}  [I(\hat{\theta})^{-1}]_{i,i}^{1/2}\}
	\end{equation}
Here $I(\hat{\theta})$ is the observd Fisher information evaluated at the MLE and  $z_{1-\alpha/2}$ is the $1-\alpha/2$ percentile of the standard Normal distribution. 
	\item[2. Likelihood-based confidence interval] 
	\begin{equation}\label{likelihood}
		\{\theta_i: \ell[\theta_i, \hat\theta_{-i}(\theta_i)] - \ell(\hat{\theta}) \geq c_{1,\alpha}/2 \},\qquad \hat\theta_{-i}(\theta_i) = \text{argmin}_{\theta_{-i}} \ell(\theta_i, \theta_{-i})
	\end{equation} 
Here $\theta_{-i}$ refers to the vector $\theta$ without the $i$th element and $(\theta_i, \hat\theta_{-i})$ is referred to as the constrained MLE.  
The $c_{1,\alpha}$ is a quantile of the Chi-squared distribution with $\Pr(\chi^2_1 \leq c_{1,\alpha})=\alpha$.
\end{description}

Note that both results are asymptotic in $n$, the dimensionality of $Y'$. P. J. Diggle and Giorgi \cite{diggle2019model} and Haskard  \cite{haskard2007anisotropic} are amongst those who propose using likelihood-based CIs for spatial models, although both note that the approximation does not hold if true value of $\theta$ is on a boundary of the parameter space, or when one or more elements of $\theta$ define the range of $Y'$. Baey, Courn`ede, and Kuhn \cite{baey2019asymptotic} and Visscher \cite{visscher2006note} show that for non-spatial models when the variances of any subset of the random effects are equal to zero, the asymptotic distribution of the test statistics is a mixture of chi-square distributions with different degrees of freedom. When true value of $\theta$ is away from the boundary the accuracy of the approximation for geostatistical models has not been investigated.

\section{Methodology}\label{methodology}

The main idea behind our method is to create likelihood-based confidence intervals by evaluating the likelihood multiple times in parallel.  We first construct a quadratic approximation to the likelihood, and then to select a large number of representative values for $\omega$ where we evaluate the likelihood.  An interpolant is built from these points and used to calculate the profile likelihood.  The time consuming part of the algorithm is evaluating the likelihood many thousands of times, and our GPU implementation performs computations for hundreds of parameter values simultaneously, with each individual likelihood evaluation further parallelized and using thousands of GPU work items.

The LGM has $7+p$ model parameters: five correlation parameters $\omega=(\phi_X,\kappa,\nu^2, \phi_{Y}, \phi_{A})$, the variance parameter $\sigma^2$, Box-Cox parameter $\lambda$, and $p$ elements of $\beta$.  Profile likelihoods for all these parameters are computed using the following steps.

\begin{enumerate}
	\setlength{\itemindent}{1.3em}
	\item [Step 1,] Obtain the MLE's $\hat{\omega}$ and $\hat{\lambda}$ by numerical maximization of the profile likelihood.
	\item [Step 2,] Use a quadratic approximation of the log-likelihood to create a set of representative parameters $\omega_1, \dots, \omega_K$ and $\lambda_1, \dots, \lambda_M$.
	\item [Step 3,] Evaluate and store various matrix determinants and cross products for the $K\cdot M$ combinations of parameters $\omega_k$ and $\lambda_k$.  Note there are only $K$ different variance matrices.
	\item [Step 4,] Compute profile likelihoods $\ell_p(\omega_k, \lambda_k)$ and obtain univariate PLLs for correlation parameters and $\lambda$ using the highest profile likelihoods.
	\item [Step 5,] Obtain PLL for multiple values of $\sigma$ by taking the maximum of\\ $\ell_p(\sigma,  \omega_k, \lambda_k,\hat{\beta}_{\sigma,\omega_k, \lambda_k};y)$ over $k$ and $m$.
	\item [Step 6,] Obtain PLL for $\beta$ by maximizing $\ell_p(\beta, \omega_k, \lambda_k,\hat{\sigma}_{\beta, \omega_k, \lambda_k};y)$ over $k$ and $m$.   Section (\ref{scalarbetaprofile}) shows how to get univariate PLL's for individual $\beta_p$ or pairs of coefficients.
\end{enumerate}

\subsection{Step 1: MLE's of covariance parameters}

Step 1 is carried out by numerically maximizing the profile likelihood $\ell_p(\omega, \lambda;y)$ in \eqref{profile}.  This requires sequential evaluations of the likelihood and is done on the CPU using the L-BFGS-B algorithm implemented in the \pkg{geostatsp} package \cite{brown2015model}.  Since the likelihood is typically very flat with respect to the shape parameter $\kappa$, two constrained maxima fixing $\kappa=0.5$ and $\kappa = 10$ are also computed.

\subsection{Step 2: A set of representative parameters}

A coarse configuration of the nuisance parameters will lead to bumps in the profile curve \cite{dauncey2015handling}, a large number of properly and sufficiently sampled parameter values would be needed to make the profile curves smooth. 
In a manner similar to the \pkg{R-INLA} software package \cite{rue2009approximate}, we create a set of ``configuration" points $\omega_k; k = 1 \ldots K$ using a quadratic approximation to the profile likelihood $\ell_p(\omega,\lambda; y)$ in \eqref{profile}.  These points distributed evenly around the surface specified by $\ell_p(\omega,\lambda; y) = c$, where $c$ takes on various $\chi_2$ quantiles from the likelihood's maximum.  If the approximation were exact, the likelihood evaluated at $\omega_k, \lambda_k$ would be equal to the $\chi^2$ quantiles of the likelihood of the model parameters. We find a reparameterization which makes exact and approximate likelihoods as similar as possible.

Step 2 involves three components: reparametrizing the likelihood function; computing a quadratic approximation using the Hessian matrix; and identifying representative points.

\noindent \textbf{\emph{Reparameterization}}

\noindent The model parameters are reparametrized in order to render the profile likelihood as close to quadratic as possible, and thereby improving the accuracy of the approximation.  The transformation of the Mat\'ern shape parameter $\kappa$ depends on whether the MLE $\hat\kappa$ is large.  The reparametrization used is
\begin{equation*}\label{eq:shapeTransform}
\tilde \kappa = \begin{cases}
	\log \kappa, & \hat\kappa < 4\\
	1/\sqrt{\kappa}, & \hat\kappa \geq 4.
\end{cases}
\end{equation*}
When $\hat\kappa$ is large the likelihood tends to be flat and the inverse-root transform proved better than the log-transform in practice.

The three parameters controlling the rate of decay in the spatial correlation, $\phi_X$, $\phi_Y$ and $\phi_A$ are reparametrized as proposed by \cite{kamal2022} with
\begin{gather*}
\gamma_1 = \log\phi_X + \log\phi_Y, \quad \text{(sumLogRange)}  \\
\gamma_2 = \sqrt{\phi_X / \phi_Y-1}\cos(2 \phi_{A}), \quad \text{(aniso1)}        \\
\gamma_3 = \sqrt{\phi_X / \phi_Y-1}\sin(2 \phi_{A}). \quad  \text{(aniso2)}
\end{gather*}
Note that $\exp(\gamma_1 /2)$ (combinedRange) is the geometric mean of $\phi_X$ and $\phi_Y$. The new parameterisation of $(\gamma_2, \gamma_3)$ applies a polar coordinate transformation to the ratio of ranges $\phi_X / \phi_Y$ and the angle $\phi_{A}$, which \cite{kamal2022} chooses because the isotropic model corresponds to the single point $\gamma_2 = \gamma_3 = 0$, 
avoiding the problem that the likelihood is flat in the $\phi_A$ dimension when data are isotropic.
The motivation for this reparameterization can be more easily understood through results presented in Section  \ref{swiss} and  \ref{soil}. 

We call the parameters $\omega = (\phi_X,\kappa,\nu^2, \phi_{Y}, \phi_{A})$ the `natural parameters', and the reparameterized parameters $\omega' = (\gamma_1, \tilde \kappa, \nu, \gamma_2, \gamma_3)$ the `internal parameters'.  

\noindent \textbf{\emph{Hessian matrix}}

\noindent There are six non-linear parameters, which are the Box-Cox parameter $\lambda$ and the five covariance parameters in $\omega'$.  For creating a quadratic approximation to the log-likelihood, we implicitly assume the second derivatives of the log likelihood with respect to $\lambda$ and $\omega'$ are zero, and focus on the 5-dimensional Hessian matrix of $\ell_p(\omega', \hat\lambda;y)$, varying $\omega'$ but fixing $\lambda$ at the MLE (as a function of $\omega'$).  The second derivative of the likelihood with respect to $\lambda$ is used to approximate the likelihood as a function of the Box-Cox parameter.  Additionally, two 4-dimensional Hessians (fixing $\kappa$ at 0.5 and 10) are computed.  

All Hessians are evaluated using numerical differentiation. For the 5-dimensional Hessian, we need to evaluate $51 \cdot 3$ likelihoods ($\hat\omega' \pm \delta$ and $\hat\lambda \pm \delta$), and for the 4-dimensional Hessian there are $33\cdot 3$ likelihoods to calculate.  The likelihoods are computed on GPU in parallel.
Flat likelihood are fairly common in spatial statistics,  

\cite{zhang2004inconsistent} shows analytically that the range, shape and variance parameters are often not well separately identifiable. It is possible that the Hessian is not negative definite or with eigenvalues very close to zero.  
We take eigen-decomposition of (minus) Hessian matrix $-H(\hat{\omega}')$ (the results will be required in obtaining representative points later):
$$-H(\hat{\omega}') = EDE^\T,$$
where $E$ is a matrix of eigenvectors and $D$ a diagonal matrix of eigenvalues, 
and replace any negative eigenvalues with their absolute values.
If the range of the eigenvalues is large, i.e., the largest one is over 100, while the smallest one is close to 0, then we replace the smallest eigenvalues 0.1, so to avoid one axes of the ellipsoid dominating the points selection.  This treatment of small and negative eigenvalues has worked well in practice for a number of real and simulated datasets.

\noindent \textbf{\emph{Obtaining representative points}}

\noindent Well-behaved log-likelihood surfaces can be well approximated around $\hat{\omega}'$ by a quadratic function derived from Taylor’s expansion
\begin{equation} \label{approximation}
\ell(\omega') = \ell(\hat{\omega}')  + \frac{1}{2} (\omega' - \hat{\omega}')^\T  (-H(\hat{\omega}'))^{-1}  (\omega' - \hat{\omega}').
\end{equation}
The solid ellipsoid of $\omega'$ values satisfying
\begin{equation} 
(\omega' - \hat{\omega}')^\T (-H(\hat{\omega}'))^{-1} (\omega' - \hat{\omega}') \leq \chi^2_p(\alpha)
\end{equation}
defines a confidence region with probability $1-\alpha$, where $\chi^2_p(\alpha)$ is the upper $(100 \alpha)$th percentile of a chi-square distribution with $p$ degrees of freedom \cite{johnson2014applied}.
In order to get points that are regularly-spaced on a hyper-ellipsoid surface, we first find points $x=(x_1, \dots x_n)$ that are uniformly spaced on a 5-dimensional unit hyper-sphere centered at the origin, by using a stochastic optimizer which maximize the distance between the closest pair of points  $\max_X \min_{i,j} \|x_i - x_j\|$. We then transform the points $x$ to $x'$ on the desired hyper-ellipsoid surface by the formula 
\begin{gather*} 
x' = E^{-2} D^{1/2} x + \hat{\omega}'.
\end{gather*}
There are 726 representative points ($n=726$) selected from each density contour level. Parameter sets from usually 9 to 11 contour levels are combined together to become the final configuration matrix for correlation parameters.  For $\lambda$, we set a vector of often 30 equally-spaced values between the 0.01 and 0.99 quantiles of its 1-dimensional quadratic approximation at $\hat\lambda$.  

Since the shape parameter is empirically difficult to estimate, a quadratic approximation may not work well on the shape parameter when its log-likelihood is far from quadratic. Hence, we  fit several additional models with the Mat\'ern shape parameter fixed at values between 0.5 and 20. The Hessian matrix is for these additional models is 4 by 4, with $n=120$ representative points from each density contour level. The likelihood is often very flat for the shape parameter, this procedure let us explore the likelihood for a range of $\kappa$ values.  It is possible to sample negative values for nugget $\nu^2$ from the quadratic approximation if the true nugget is close to 0, in that case, we turn half of the negative values to 0, and the other half to random uniform numbers sampled from the interval $(0,2)$, see Figure \ref{fig:m3} for a plot of the sampled nugget values and their corresponding profile log-likelihoods in a real data example, in which a visible small pile of points is collected at 0.

\subsection{Step 3: Log-likelihood computation}

The computationally expensive steps in evaluating the likelihood are taking the Cholesky decomposition of the variance matrix $V$ and performing  triangular backsolve computations.  These steps are performed on the GPU, and the crossproducts and matrix determinants listed in Table \ref{tab:tocpu} are saved and returned to the CPU.  These components of the likelihoods are then used to compute different variations of the likelihood (i.e.\ profiling out $\beta$ or $\sigma$ or both).

\begin{table}[tb]
	\centering
	\makebox[\textwidth]{
	\begin{tabular}{lll}
	\toprule  
		Component & name in {\tt R} & \\
		\hline
		$\log|V|$ & {\tt detVar} & \\
		$\log|X^\T V^{-1} X|$ & {\tt detReml}\\
		$y^\T V^{-1} y$ & {\tt ssqYX} & upper-left diagonal entries \\
		$X^\T V^{-1} X$ & {\tt ssqYX } & lower-right diagonal entries\\
		$X^\T V^{-1}y$ & {\tt ssqYX} & lower-left block \\
		$(X \hat\beta)^\T V^{-1}(X \hat\beta)$ & {\tt ssqBetahat}\\
		$(y-X\hat{\beta})^\T V^{-1}(y-X\hat{\beta})$ & {\tt ssqResidual}\\
		\bottomrule
	\end{tabular}}
	\caption{Summary statistics computed and retained for each set of parameter values}
	\label{tab:tocpu}
\end{table}

We take the following steps to compute the items in Table \ref{tab:tocpu} for the parameter sets $\omega'_1$ to $\omega'_K$.
\begin{enumerate}
	\setlength{\itemindent}{1.3em}
	\item[Step 1:] Compute the Mat\'ern correlation matrices $V=(V_1, \dots, V_K)$ with the \fct{maternBatch} function.  
	\item[Step 2:] Take the Cholesky decompositions $V_k=L_kD_kL_k^\T$ with \fct{cholBatch}, where the $L_k$ are lower unit triangular matrices and the $D_k$ are the diagonal matrices.  The function, then $\log|V|=2\log|L|+\log|D|=\log|D|$. \fct{cholBatch} returns $L_k$, $D_k$ and $\log|V_k| = \text{sum}(\log D_k)$.
	\item[Step 3:] Solve $L_k^{-1}(y,X)$ with \fct{backsolveBatch}.
	\item[Step 4:] Use the above to compute $(y, X)^\T V_k^{-1} (y,X)$ with \fct{crossprodBatch}, the results are saved in {\tt ssqYX}. 
	\item[Step 5:] Perform a Cholesky decomposition on $X_k^\T V_k^{-1} X_k=Q_kP_kQ_k^\T$ with \fct{cholBatch} and calculate $\log|X_k^\T V_k^{-1} X_k|$.
	\item[Step 6:] Solve the equation $c=Q_k^{-1}X^\T V_k^{-1}y$ with \fct{backsolveBatch}.
	\item[Step 7:] Compute $\text{\tt ssqBetahat} = (X \hat\beta)^\T V^{-1}(X \hat\beta)=c^\T P_k^{-1}c$ by \fct{crossprodBatch}.
	\item[Step 8:] Compute $\text{\tt ssqResidual}=(y-X\hat{\beta})^\T V^{-1}(y-X\hat{\beta})=\text{\tt ssqYX} - \text{\tt ssqBetahat}$. 
\end{enumerate} 
GPU memory limitations mean not all $K$ of the parameter sets can be processed simultaneously, the limiting factor is the memory occupied by the variance matrices $V_k$.  The results presented here process several hundred parameter sets at a time, the maximum number possible depends on the size of the dataset and the amount of GPU memory.

\subsection{Profile likelihood for $\beta_p$} \label{scalarbetaprofile}

Here we consider computing the profile likelihood for a single regression coefficient $\beta_p$ for the $p$th covariate.  The profile likelihood is
\begin{align} \label{profilebetai}
	\ell_p(\beta_p; y) &= 
	\underset{ \beta_{-p}, \sigma, \omega, \lambda}{\text{max}} 
	\{\ell(\beta_p, \beta_{-p}, \sigma, \omega, \lambda;y)\} \notag \\
	&\approx
	\underset{k,m}{\text{max}}\{
	\ell (\beta_p, \hat\beta_{-p}(\beta_p, \omega_k, \lambda_\ell),
	\hat\sigma(\beta_p, \omega_k, \lambda_\ell), 
	\omega_k, \lambda_m)
	\}
\end{align}
where $\beta_{-p}$ is the vector $\beta$ without its $p$th element.  The functions $\hat\sigma(\cdot)$ and $\hat\beta_{-p}(\cdot)$ denote the the MLE's of $\beta_{-p}$ and $\sigma$ for given values of $\beta_p$, $\omega$ and $\lambda$, for which there are closed-form solutions.

When $\beta_p$ is fixed, the term $X_p\beta_p$ becomes a fixed offset and the term in the model involving covariates becomes $X_{-p}\T\beta_{-p}$, where $X_{-p}$ is the matrix $X$ without the $p$th column.  Computing $\hat\beta_{-p}(\beta_p, \omega, \lambda)$ involves the following matrix crossproducts:
\begin{align*}
	(y - X_p \beta_p)^\T V^{-1}(y - X_p \beta_p)  &= y^\T V^{-1}y - 2 \beta_p (X^\T V^{-1} y)_{[p,]} + \beta_p^2 (X^\T V^{-1} X)_{[p,p]},\\
	(X_{-p}\hat{\beta_{-p}})^\T V^{-1} (X_{-p}\hat{\beta_{-p}}) 
	&= (X^\T V^{-1}y)^\T_{[-p,]} (X^\T V^{-1} X)^{-1}_{[-p,-p]} (X^\T V^{-1}y)_{[-p,]}\\
	&- 2\beta_p (X^\T V^{-1}X)^\T_{[-p, p]} (X^\T V^{-1} X)^{-1}_{[-p,-p]} (X^\T V^{-1}y)_{[-p,]}\\
	&+ \beta_p^2 (X^\T V^{-1}X)^\T_{[-p, p]} (X^\T V^{-1} X)^{-1}_{[-p,-p]} (X^\T V^{-1}X)_{[-p,p]}.
\end{align*}
The subscript notation $[-p, -p]$ indicates the $p$th row and column is excluded from a matrix, with $[-p, p]$ refers to the $p$th column of a matrix while excluding the $p$th row.  Computing the above only needs subsetting rows and columns from the {\tt ssqYX} matrices listed in Table \ref{tab:tocpu}, and the profile likelihoods for each $\beta_p$ can be computed without additional use of the GPU.

\subsection{Profile likelihood for $\sigma$}

By substituting \eqref{betahat} into \eqref{eq:2}, it is easy to obtain the PLL for $(\sigma, \omega, \lambda)$ as
\begin{align} \label{profileSigma}
	-2\ell_p(\sigma,\omega, \lambda;y) - n\log(2\pi) =&(y-X\hat{\beta})^\T (\sigma^2 V)^{-1}(y-X\hat\beta)+ n \log\sigma^2     \notag\\
	& +\log|V| -2(\lambda-1)\sum_{i=1}^{n}\log y_i,
\end{align}
where $\hat\beta$ and $V$ both depend on $\omega$.  
The various terms in (\ref{profileSigma}) can all be computed with the quantities in Table \ref{tab:tocpu} and the exception of the final term (the Box-Cox Jacobian) which is also returned by the GPU.
We approximate the PLL for $\sigma$ with
$$
\ell_p(\sigma) \approx \max_{k,m} \ell_p(\sigma,\omega_k, \lambda_m ;y),
$$  
and evaluate $\ell_p(\sigma)$ for a range of values of $\sigma$.

\subsection{Profile likelihood for correlation parameters}

We obtain the profile likelihood for any of the  correlation parameters, either reparametrized $\omega' = (\gamma_1, \tilde \kappa, \nu, \gamma_2, \gamma_3)$ or original $\omega = (\phi_{X},\kappa,\nu^2, \phi_{Y}, \phi_{A})$, from the points cloud of all computed likelihoods $\ell_p(\omega_k, \lambda_m;y)$ like the example plots shown in Figure \ref{swissprofile1}.  To compute the PLL for the shape parameter $\kappa$, for example, we create a 2D scatterplot from the configuration points $\omega_k$ with $\kappa_k$ on the x-axis and $\underset{ m}{\text{max}} \ell_p(\omega_k, \lambda_m;y)$ on the y-axis.  The collection of `best' likelihoods are obtained by computing the convex hull of these points, shown as the 12 black circular points in Figure \ref{fig:m2}. The green line in Figure \ref{fig:m2}, a linear interpolation of these 12 points, is the estimated PLL curve.  The PLL for the Box-Cox parameter $\lambda$ is simply the interpolation of $\lambda_m$ and $\underset{ k}{\text{max}}  \ell_p(\omega_k, \lambda_m;y)$.

Similarly, for 2-dimensional profile likelihood of any 2 of the correlation parameters $(\omega'_i, \omega'_j)$ in $\omega'$, we first find the smallest convex hull that encloses the set of points $[\omega'_i, \omega'_j, \underset{ \lambda}{\text{max}} \{\ell_p(\omega, \lambda;y)\}]$ in the 3-dimensional space, remove points on the bottom facet of the hull, and then the profile likelihood surface is linearly interpolated. We create regular girds on the space spanned by $(\omega'_i, \omega'_j)$, and predict values for these grids using the interpolation.

For plotting 2-dimensional profile likelihoods of the correlation parameters in the ``natural space" $(\omega_i, \omega_j)$, we convert the regular girds created in ``natural space" to grids in ``internal space", and prediction for those grid points still uses interpolation formed in the ``internal space". The convex hull and interpolation method works poorly in the original parameter space, since the log likelihood is less bell-shaped than in the reparametrized space.

\section{Computational considerations}
Taking full advantage of the parallel computing abilities of GPU's requires some manual configuration of the GPU memory and work items (the GPU equivalent of CPU cores).  Work items are divided into work groups which share local memory (which is faster than global memory).  We treat work items as being on a two-dimensional grid (a third dimension is unused), and divide tasks amongst work items as shown in Table \ref{tab:gpuFunction}.

Work items are arrayed on an \code{Nglobal0} by \code{Nglobal1} grid, with each work group containing \code{Nlocal0} by \code{Nlocal1} work items.
Suppose \code{Nglobal0=6}, \code{Nglobal1=4}, \code{Nlocal0=3} and \code{Nlocal1=2}, which means we have $2\times2$ work-groups on the 2-dimensional (dimension 0 and dimension 1) computation domain (See Figure \ref{fig:gpupic}), each work-group consists of $3\times 2$ work-items. Suppose we have a total of 500 parameter configurations divided into 5 sets of 100. With each looping inside the \proglang{\texttt{C++}} function \fct{likfitGpu\_Backend}, the GPU computes 100 log-likelihoods and saves corresponding 100 sufficient statistics on the GPU's global memory.

\begin{table}[h]
\centering
  \caption{List of the main GPU kernels showing the tasks they compute and how the task is divided amongst work items and work groups.}
  \makebox[\textwidth]{
	$\begin{array}{ lccccc }
	\toprule
		\text{Kernel}  & \text{Formula} & \text{group0} & \text{group1} & \text{local0}  &\text{local1}\\
		\midrule
		\text{maternBatch} &V         &\text{shared coordinates}     & \text{shared parameters}   & \text{matrix element}  & \text{matrix}\\
		\text{cholBatch} &V=LDL^\T  &\text{matrix}  & unused & \text{row of L}         & \text{column of L} \\
		\text{backsolveBatch} &AC=B &\text{matrix} & \text{col of B and C} & \text{row of A and C} & \text{col of A, row of C} \\
		\text{crossprodBatch}    &C=A^\T DA   & \text{column of C}   & \text{matrix}  & \text{elements of D}  & \text{row of C} \\
		\bottomrule
	\end{array}$
	}
	\label{tab:gpuFunction}
\end{table}

Table \ref{tab:gpuFunction} shows the four main GPU kernels which are used to perform matrix operations. In each of the loop in this example, the \emph{maternBatch} kernel has work items in dimension 0 looping through elements of the variance matrix and work items in dimension 1 looping through parameter sets (each one used for a different variance matrix).  In the scenario above each work group will use half of the 100 parameter sets and copy 50 sets of parameters to local memory, since there are two work groups in dimension 1.
Each work-group copies \code{NparamPerIter}$/2=50$ parameters and 3 (the first element of \code{Nlocal}) coordinates in the local memory.  At each iteration a work group will copy over pairs of coordinates for three matrix elements to local memory (since {\tt Nlocal0=3}), and use these coordinates for two matrices (since \code{Nlocal1 = 2}).

The Cholesky decomposition is performed by \emph{cholBatch} and only uses a single work group in dimension 1 (requiring {\tt Nglobal1} = {\tt Nlocal1}). The first work-group takes Cholesky decomposition for variance matrices $V_0, V_2, V_4, \dots$, the second work-group takes cholesky decomposition for $V_1, V_3, V_5, \dots$. Within each work-group, the first-row work-items are responsible for the $1,4,7,\dots$ rows of the first column of $L$, and simultaneously, the second-row work-items is responsible for the $2,5,8,\dots$ rows of the first column of $L$, the third-row calculates cells in the $3,6,9,\dots$ rows of the first column of $L$.

In the \emph{backsolveBatch} kernel, work-groups of the same row computes the same $C_i$'s, in the example here, the upper two work-groups compute $C_0, C_2, C_4, \dots$ in order, and the lower two work-groups computes the rest $C_1, C_3, C_5, \dots$ in sequence. Each work-group is responsible for \code{ceil(ncol(B)/2)} columns of B, in other words, work-groups shaded grey color here are responsible for the columns $0,2,4,\dots$ of each $C_i$, and the white work-groups are responsible for columns $1,3,5,\dots$ of each $C_i$. Within each work-group, work-items on row 0 computes each of the rows $0,3,6,\dots$ of each $C_i$ together, similarly and simultaneously, work-items on row 1 computes each of the rows $1,4,7,\dots$ of each $C_i$ together, and work-items on row 2 computes each of the rows $2,5,8,\dots$ of each $C_i$ together. 

The \emph{crossprodBatch} kernel computes $C=A^\T DA$ or $C=A^\T D^{-1}A$ or $C=A^\T A$. The two shaded work-groups on the left are responsible for computing $C_O, C_2, C_4, \dots$, and the white work-groups are responsible for computing $C_1, C_3, C_5, \dots$. The group index in dimension 0 decides the columns of C that the work-items in this work-group compute for, here the upper work-group computes for the columns $0,2,4,\dots$ in each $C_i$, and the lower work-group computes the columns $1,3,5,\dots$ of each $C_i$. Within each work-group, work-items of same local index in dimension 1 compute the same rows of $C_i$, so here the left three work-items compute rows $0,2,4,\dots$ of each $C_i$ and the right three work-items compute rows $1,3,5,\dots$ of each $C_i$.
\begin{knitrout}
	\definecolor{shadecolor}{rgb}{1, 1, 1}\color{fgcolor}
	\begin{figure}[H]
		\centering 
			\includegraphics[width=0.42\linewidth]{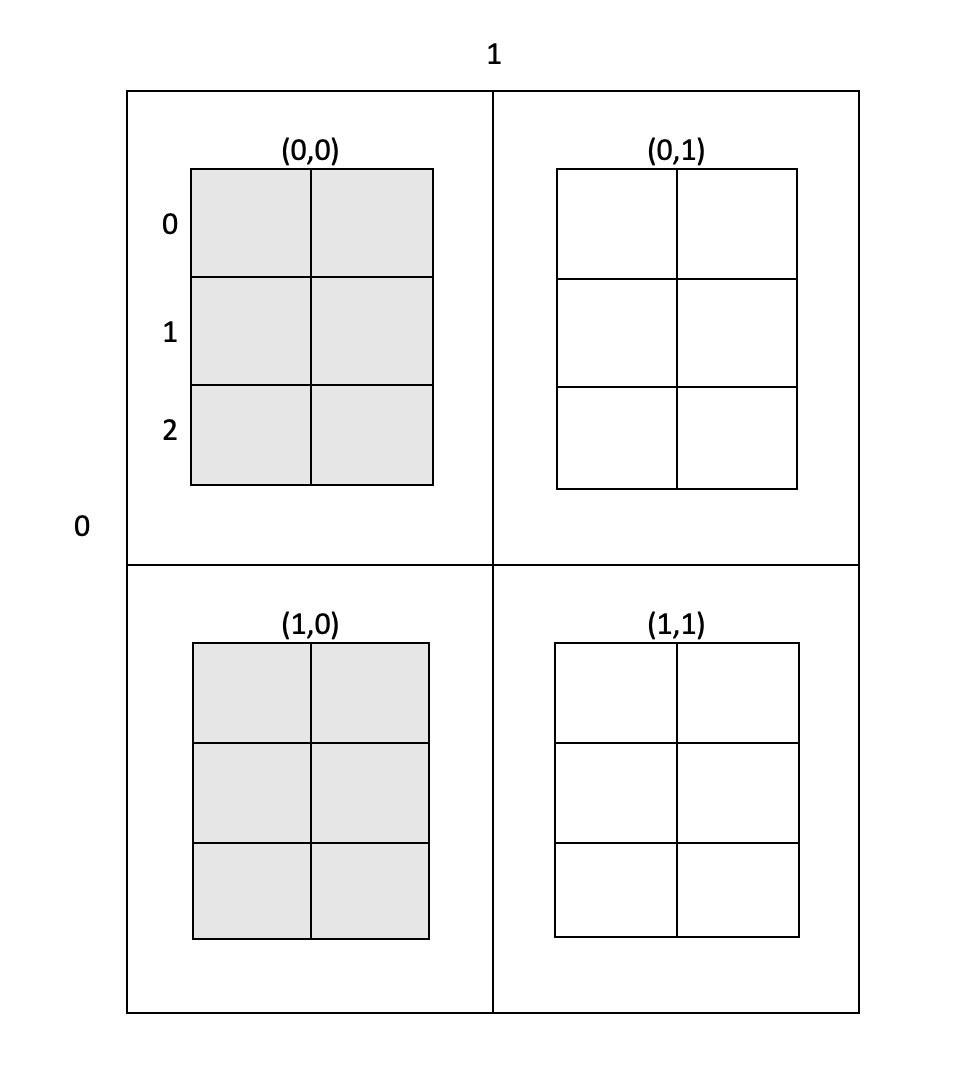} 			
		\caption[Representation of 2-dimensional computation domain for the simple example for illustration]{Representation of 2-dimensional computation domain for the simple example for illustration.}\label{fig:gpupic}
	\end{figure}	
\end{knitrout}
There is the argument \code{NparamPerIter} in the function \fct{likfitLgmGpu} which specifies how many parameter sets are simultaneously computed on each iteration of calls to the back-end GPU functions. The batch size  value (\code{NparamPerIter}) is limited by the available GPU global memory. If the LGM to estimate has $n$ observations, $p$ covariates, and the vector configuration for the Box-Cox parameter has $m$ values. Then each loop inside the backend \proglang{\texttt{C++}} function refreshes the GPU matrix $V$ (stores Mat\'ern correlation matrices), which has $n$*\code{NparamPerIter} rows and $n$ columns; \emph{crossprodBatch}'s result \code{ssqYX} is a GPU matrix of $(m+p)$*\code{NparamPerIter} rows and $m+p$ columns.  We should make \code{NparamPerIter} relatively small if it is a large data set to fit or if there are lots of covariates in the model and configuration for $\lambda$ has a large number of values. The values of \code{Nglobal} and \code{Nlocal} are better to be powers of two, and should not exceed the maximum number of work-items on the GPU, which can be found by \code{gpuR::gpuInfo()}.

\section{Two simulation studies}

To assess the CIs produced by our method in a realistic setting, we performed 2 simulation studies, study A uses anisotropic data and study B uses isotropic data. Within each study, we do 150 simulations using simulated coordinates, and the other 150 simulations using coordinates from the classic \emph{Loaloa} dataset.

Specifically, in study A, we generate 150 realizations $y_1, \dots, y_{150}$ twice from the model $Y_i \sim N(5 + 1*X_{i1} + 1*X_{i2} + U(s_i), \tau^2)$,
where $\tau = 0.8$.
In the two simulations A(i) and A(ii), the stationary Gaussian random field $U(s)$ with variance $\sigma^2 = 1$ and Mat\'ern correlation function with parameters $\kappa = 2, \phi_R = 2, \phi_A = 0.2$ is produced over different coordinates, and the two explanatory variables are defined in different ways accordingly. No Box-Cox transformation was performed on $Y$ so the parameter $\lambda = 1$. 

For study A(i), $U(x)$ is generated over 224 (Ngrid = 15) point locations on a $9km \times 9km$ square (displayed in Figure \ref{fig:sim1coords-1}), the point locations in Figure \ref{fig:sim1coords-1} are randomly generated, we removed one point that was very close to another point. The range parameter $\phi_X$ is set to be 1,000 for this coordinate.  Covariates $X_{i1}$ and $X_{i2}$ were defined using the coordinate values: $X_{i1} = coordsX_i/10000$, $X_{i2} = (coordsX_i/10000)^2$, where $coordsX_i$ represents the x coordinate of location $s_i$.

For study A(ii), $U(s)$ is generated on the \emph{Loaloa} coordinates, shown in Figure \ref{fig:sim1coords-2}. \emph{Loaloa} is a real data set (available in the \pkg{geostatsp} package) that has 190 locations (from 190 village surveys), many of them are clustered together. For A(ii), $\phi_X$ is set as 50,000. Correspondingly, the two explanatory variables were defined as: $X_{i1} = \log(elevationLoa_i); X_{i2} = eviLoa_i*10^{-7}$,  where $elevationLoa_i$ means the log of elevation at location $s_i$.

Similarly, for study B, we produce a total of 300 realizations $y_1, \dots, y_{300}$ from the model $Y_i \sim N(2 + 1*X_{i1} + 1*X_{i2} + U(s_i), \tau^2)$
with $\tau = 0.8$ and $U(s)$ defined on the same two types of coordinates as in study A. $U(s)$ have the same parameter setup as in simulation A, except for $U(s)$ being isotropic. Again, study B(i) has realizations produced on the simulated coordinates, study B(ii) uses the \emph{Loaloa} coordinates.
\begin{knitrout}
	\definecolor{shadecolor}{rgb}{1, 1, 1}\color{fgcolor}
	\begin{figure}[H]
		
		{\centering \subfloat[Simulated\label{fig:sim1coords-1}]{\includegraphics[width=0.47\linewidth]{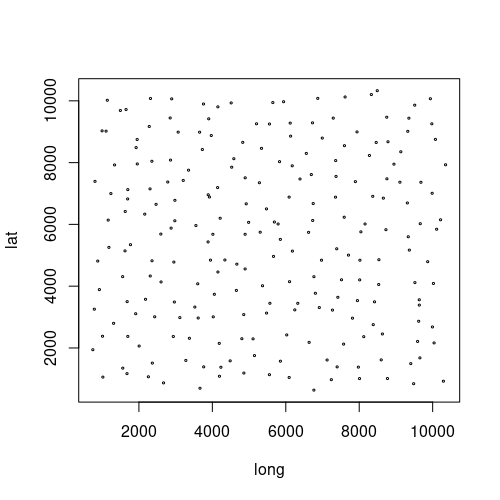} }
			\subfloat[Loaloa\label{fig:sim1coords-2}]{\includegraphics[width=0.47\linewidth]{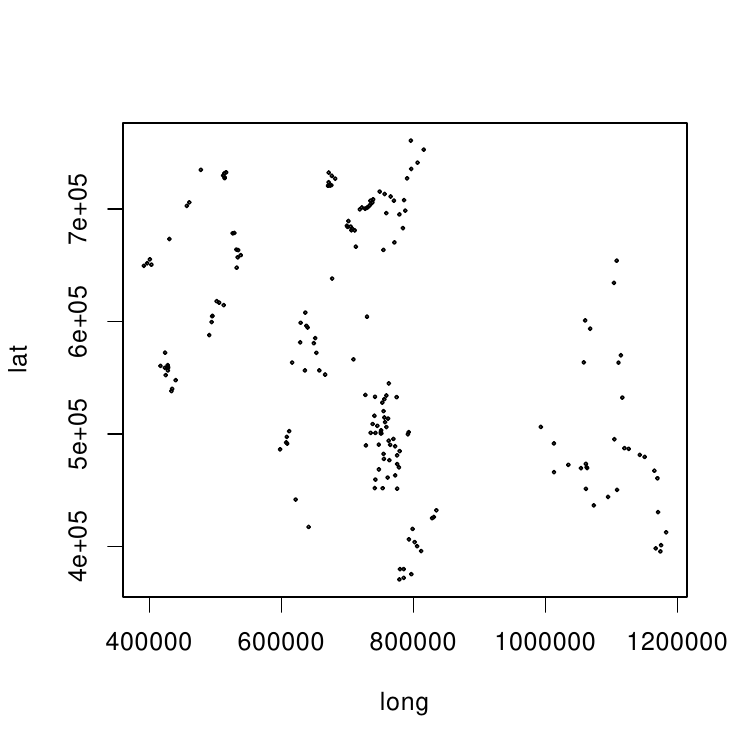} }
			
		}		
		\caption{\label{fig:sim1coords}Simulation  coordinates for $U(s)$.}
	\end{figure}
	
\end{knitrout}

\begin{figure}[H]
	\centering
	\subcaptionbox{$\log(\phi_X \phi_Y)$}{\includegraphics[width=0.44\textwidth]{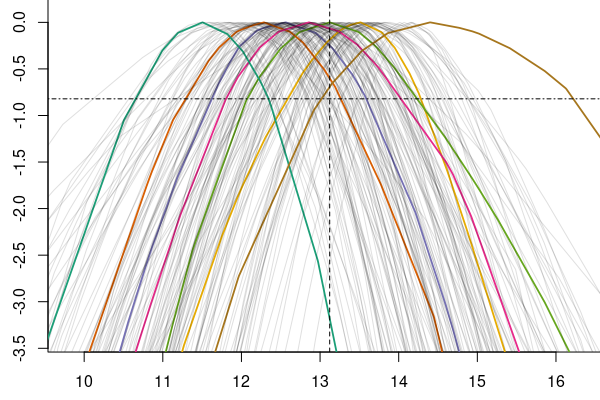}}
	\subcaptionbox{$\kappa$}{\includegraphics[width=0.44\textwidth]{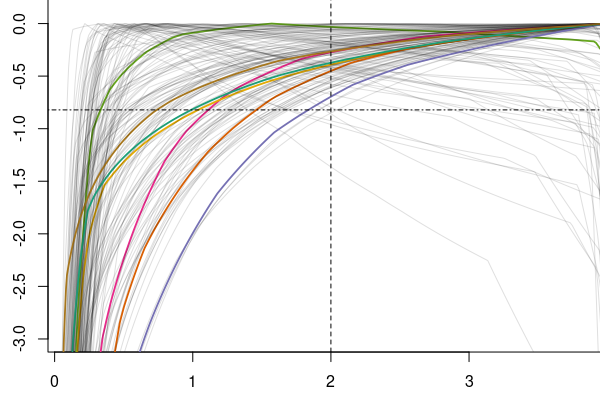}}
	\subcaptionbox{$\nu^2$}{\includegraphics[width=0.44\textwidth]{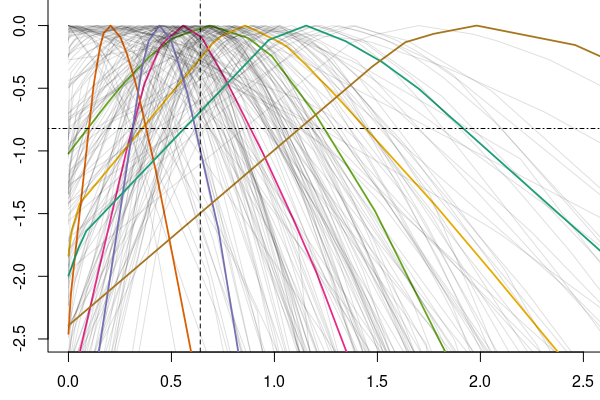}}
	\subcaptionbox{$\gamma_2$}{\includegraphics[width=0.44\textwidth]{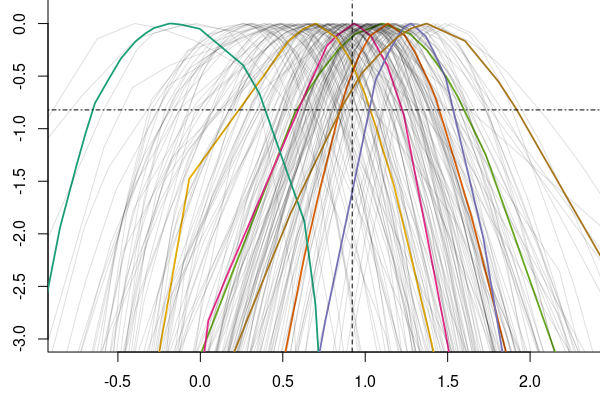}}
	\subcaptionbox{$\gamma_3$}{\includegraphics[width=0.44\textwidth]{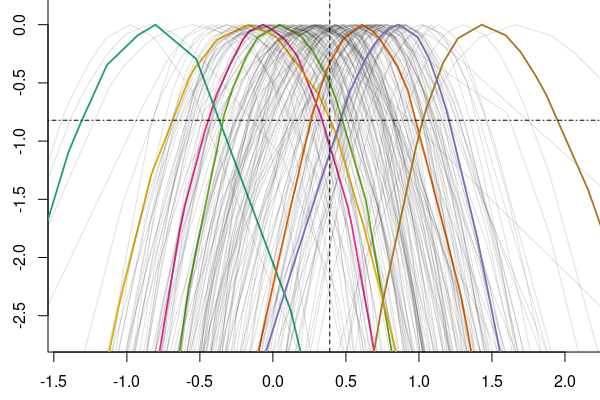}}
	\subcaptionbox{Intercept}{\includegraphics[width=0.44\textwidth]{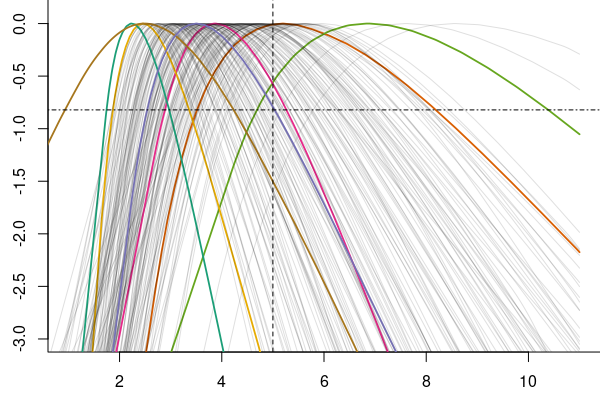}}
	\subcaptionbox{$\beta_1$}{\includegraphics[width=0.44\textwidth]{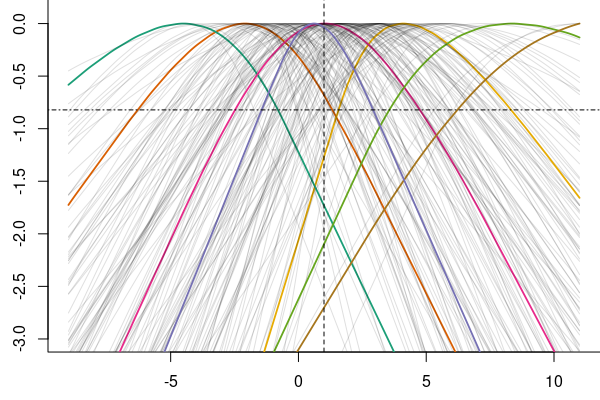}}
	\subcaptionbox{$\beta_2$}{\includegraphics[width=0.44\textwidth]{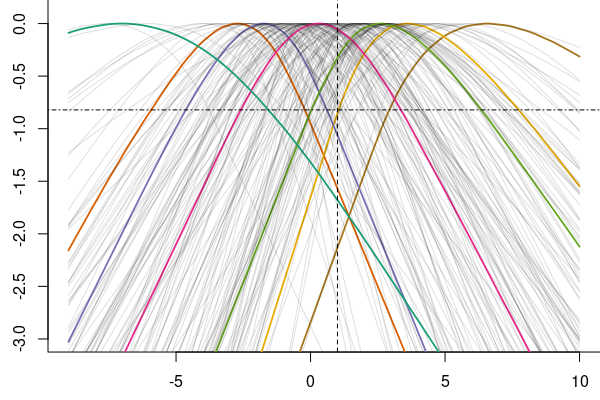}}
	\caption{Profile log-likelihoods plots for the model parameters in simulation Study A (i) (Anisotropic data and simulated coordinates). }\label{fig:sim1graphs}
\end{figure}

\begin{figure}[H]
	\centering
	\subcaptionbox{$\log(\phi_X \phi_Y)$}{\includegraphics[width=0.44\textwidth]{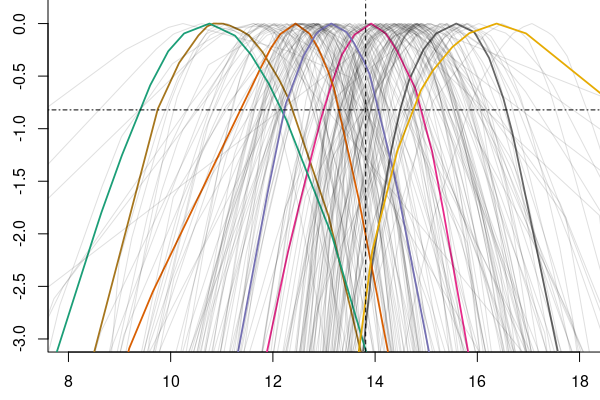}}
	\subcaptionbox{$\kappa$}{\includegraphics[width=0.44\textwidth]{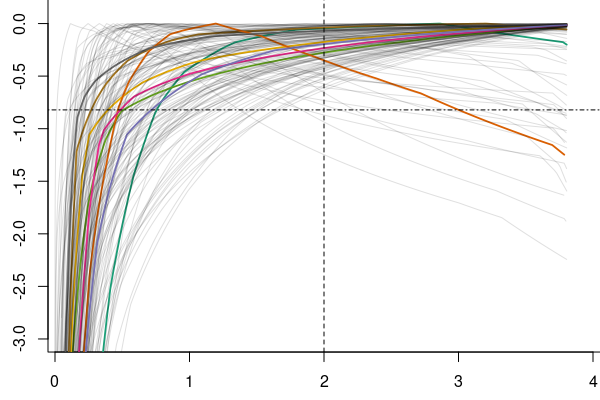}}
	\subcaptionbox{$\nu^2$}{\includegraphics[width=0.44\textwidth]{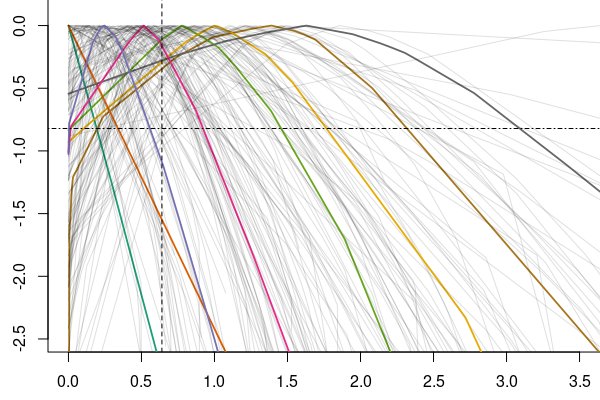}}
	\subcaptionbox{aniso1}{\includegraphics[width=0.44\textwidth]{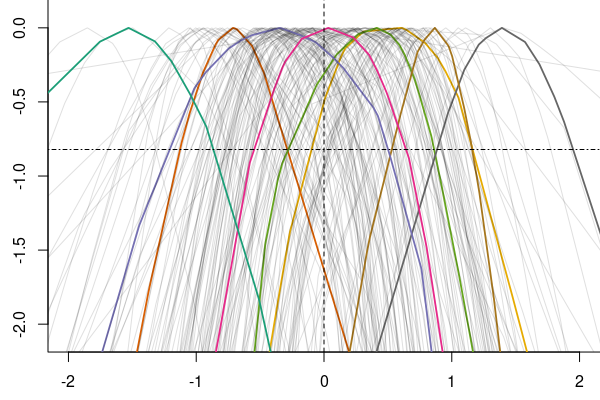}}
	\subcaptionbox{aniso2}{\includegraphics[width=0.44\textwidth]{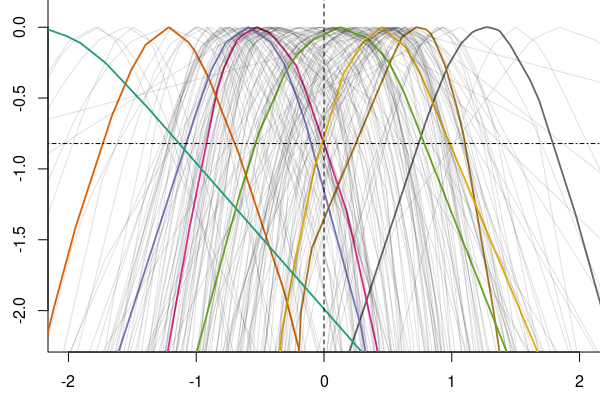}}
	\subcaptionbox{Intercept}{\includegraphics[width=0.44\textwidth]{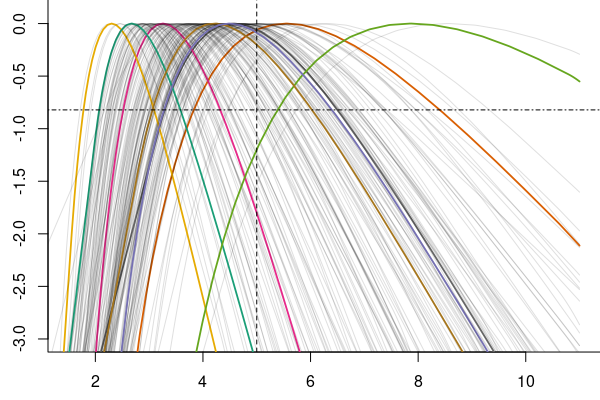}}
	\subcaptionbox{$\beta_1$}{\includegraphics[width=0.44\textwidth]{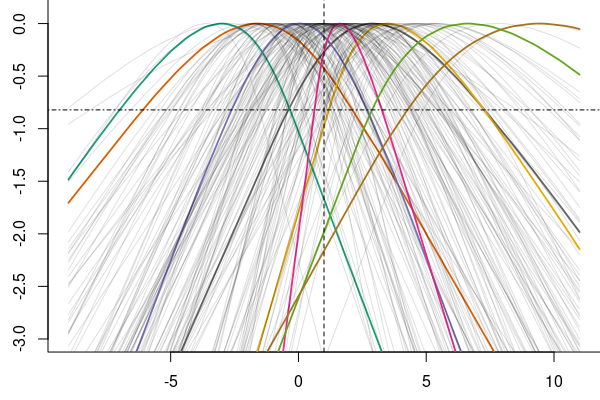}}
	\subcaptionbox{$\beta_2$}{\includegraphics[width=0.44\textwidth]{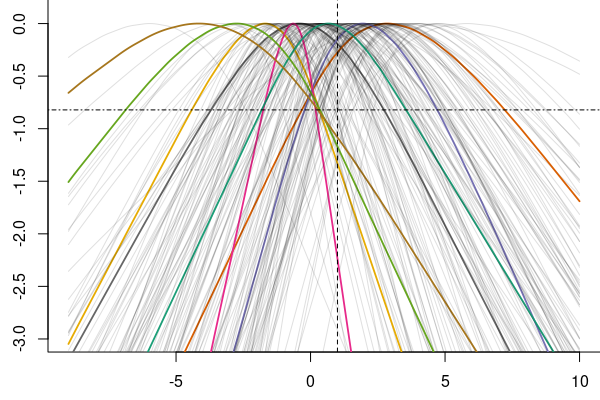}}
	\caption{Profile log-likelihoods for the model parameters in simulation Study B (i) (Isotropic data and simulated coordinates). }\label{fig:sim2graphs}
\end{figure}

\begin{figure}[H]
	\centering
	\subcaptionbox{$\gamma_2$ vs $\gamma_3$}{\includegraphics[width=0.49\textwidth]{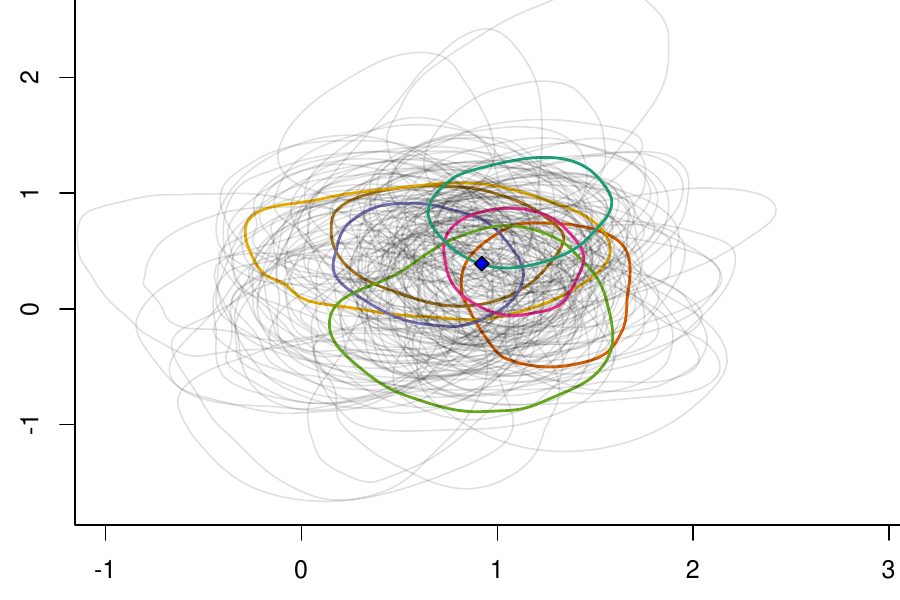}}
	\subcaptionbox{$\gamma_2$ vs $\gamma_3$}{\includegraphics[width=0.49\textwidth]{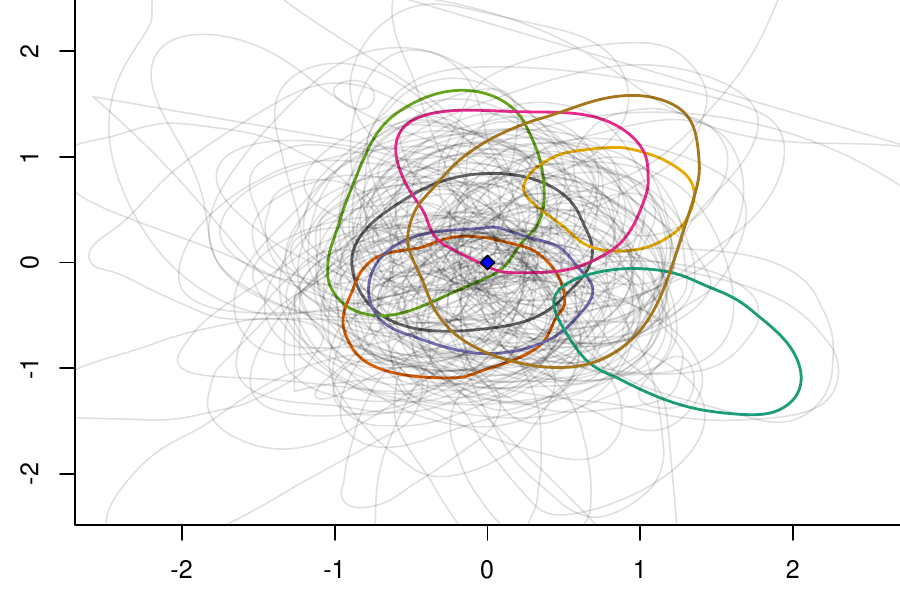}}
	\caption{Plot of the 2-dimensional 80\% confidence regions of the profile log-likelihoods for $\gamma_2$ (on the x axis) and $\gamma_3$ (on the y axis) from the 150 simulations in Study A (i) (left panel) using anisotropic data and regular coordinates and B (i) (right panel) using isotropic data and regular coordinates, respectively. }\label{fig:simgraphs2d}
\end{figure}

We computed coverage for 80\% CIs for each of the 8 parameters 
and compared the CI coverage rates for these parameters with results computed using \pkg{geostatsp} in Table \ref{summarycompare}. The 150 PLL curves using simulated coordinates for the 8 parameters in study A(i) and B(i) are shown in Figure \ref{fig:sim1graphs} and \ref{fig:sim2graphs} respectively. 2-dimensional plots of the 80\% confidence regions of the anisotropic parameters are shown in Figure  \ref{fig:simgraphs2d}. 

The results demonstrate that \pkg{gpuLik} (or profile-likelihood-based CIs) have higher and closer to expected coverage than \pkg{geostatsp}'s (or information-based CIs). The superiority is much more obvious for the regression parameter $\beta$'s and for simulation based on the real coordinates \emph{Loaloa}.

For simulations using isotropic data, we also counted the coverage of the 2-dimensional 80\% profile likelihood-based confidence regions of $(\gamma_2,\gamma_3)$, which are shown in Figure \ref{fig:simgraphs2d}. Out of 150 simulations, 103 confidence regions covered the true values of the parameters $(\gamma_2,\gamma_3)$ (rate $= 0.687$).

\begin{knitrout}
\definecolor{shadecolor}{rgb}{0.969, 0.969, 0.969}\color{fgcolor}
\begin{table}[H]
\caption{\label{summarycompare}A comparison between the coverage rates of the 150 80\% confidence intervals of the model parameters in simulation studies A and B using methods provided by \pkg{gpuLik} and \pkg{geostatsp} respectively.}
\centering
\makebox[\textwidth]{
\begin{tabular}[t]{lrrrrrrrr}
\toprule
\multicolumn{1}{c}{ } & \multicolumn{4}{c}{Anisotropic} & \multicolumn{4}{c}{Isotropic} \\
\cmidrule(l{3pt}r{3pt}){2-5} \cmidrule(l{3pt}r{3pt}){6-9}
\multicolumn{1}{c}{ } & \multicolumn{2}{c}{Simulated coords} & \multicolumn{2}{c}{Loaloa coords} & \multicolumn{2}{c}{Simulated coords} & \multicolumn{2}{c}{Loaloa coords} \\
\cmidrule(l{3pt}r{3pt}){2-3} \cmidrule(l{3pt}r{3pt}){4-5} \cmidrule(l{3pt}r{3pt}){6-7} \cmidrule(l{3pt}r{3pt}){8-9}
Parameters & likelihood-based & Wald & likelihood-based & Wald & likelihood-based & Wald & likelihood-based & Wald\\
\midrule
Intercept & 0.600 & 0.340 & 0.740 & 0.513 & 0.666 & 0.340 & 0.686 & 0.533\\
$\beta_1$ & 0.800 & 0.786 & 0.813 & 0.173 & 0.800 & 0.786 & 0.813 & 0.180\\
$\beta_2$ & 0.800 & 0.766 & 0.820 & 0.180 & 0.793 & 0.746 & 0.813 & 0.266\\
$\phi_X$ or $\sqrt{\phi_X \phi_Y}$ & 0.760 & 0.460 & 0.760 & 0.706 & 0.420 & 0.193 & 0.513 & 0.400\\
$\kappa$ & 0.967 & 0.800 & 0.926 & 0.766 & 0.980 & 0.393 & 0.933 & 0.506\\
$\nu^2$ or $\tau$ & 0.860 & 0.780 & 0.826 & 0.713 & 0.753 & 0.320 & 0.826 & 0.466\\
$\gamma_2$ or $\phi_R$ & 0.780 & 0.713 & 0.660 & 0.580 & 0.653 & 0.186 & 0.660 & 0.320\\
$\gamma_3$ or $\phi_A$ & 0.713 & 0.680 & 0.686 & 0.573 & 0.680 & 0.213 & 0.733 & 0.353\\
$\lambda$ & 0.806 & 0.720 & 0.793 & 0.686 & 0.813 & 0.326 & 0.820 & 0.480\\
\bottomrule
\end{tabular}
}
\end{table}
\end{knitrout}

\section{Example: Swiss rainfall data} \label{swiss}

\begin{figure}[tb]
		{\centering \includegraphics[width=0.6\linewidth]{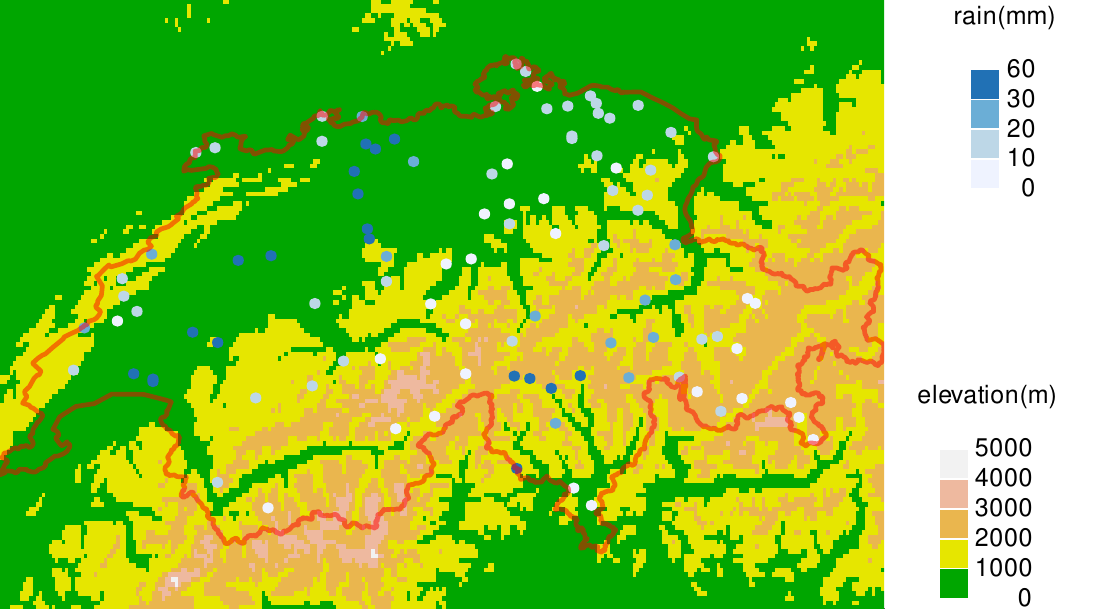} 
			
		}		
		\caption[Swiss rainfall data (colored blue points and top legend) with elevation (background colors and bottom legend)]{Swiss rainfall data (colored blue points and top legend) with elevation (background colors and bottom legend).}\label{swissmap}
\end{figure}

We now implement our method on the Swiss rainfall data from the \pkg{geostatsp} package  (See \cite{brown2015model,brown2021package} for a detailed description of the data and the SIC-97 project).  
The data, plotted in Figure \ref{swissmap}, consists of 100 daily rainfall measurements made in Switzerland in May, 1986.
We fit the data with the Linear Geostatistical Model
\begin{gather*}\label{swissmodel}
	Y'_i \sim \text{N} [ X(s_i)^\T  \beta  + U(s_i), \tau^2 ]\\
	\text{COV}[U(s+h), U(s)] = \sigma^2 \rho(h;\phi_X, \phi_Y, \phi_A, \kappa) 
\end{gather*}
where $Y'_i$ is the (Box-Cox transformed) rainfall measurement at the $i$th location, the covariates $X(s)$ are comprised of an intercept and the elevation at location $s$, and $U(s)$ is a zero-mean Gaussian process.  The correlation function $\rho$ is an anisotropic Mat\'ern with range parameters $\phi_X$ and $\phi_Y$, a rotation parameter $\phi_A$, and shape parameter $\kappa$.  
Below the LGM is fit using \code{swissRain} and \code{swissAltitude} objects from \pkg{geostatsp}, with the \fct{geostatsp::lgm} function computing the MLE's of the model parameters.
\begin{knitrout}
\definecolor{shadecolor}{rgb}{1, 1, 1}\color{fgcolor}\begin{kframe}
\begin{verbatim}
R> data("swissRain", package = "geostatsp")
R> swissRain$elevation = raster::extract(swissAltitude, swissRain)
R> mle1 = geostatsp::lgm(formula = rain ~ elevation, data = swissRain,
+      covariates = swissAltitude, fixBoxcox = FALSE, fixShape = FALSE,
+      fixNugget = FALSE, aniso = TRUE, reml = FALSE, grid = 20)
\end{verbatim}
\end{kframe}
\end{knitrout}
As described in Step 1 of the algorithm in Section \ref{methodology}, we fit five additional models fixing the Mat\'ern shape parameter $\kappa$ at 0.5, 0.9, 10, 20, and 100 (using arguments \code{fixNuggeet = FALSE} and \code{shape = 0.5}).  A list containing these six outputs from \fct{lgm} is passed to the \fct{gpuLik::configParams} function to create representative points. 
\begin{knitrout}
\definecolor{shadecolor}{rgb}{1, 1, 1}\color{fgcolor}\begin{kframe}
\begin{verbatim}
R> reParamaters = gpuLik::configParams(list(mle1, mle2, mle3, mle4, mle5,
+      mle6), alpha = c(0.00001, 0.01, 0.1, 0.2, 0.25, 0.3, 0.5, 0.8,
+      0.9, 0.95, 0.99, 0.999))
R> paramsUse <- reParamaters$representativeParamaters
R> b <- reParamaters$boxcox
\end{verbatim}
\end{kframe}
\end{knitrout}
We select 12 significance levels for $\alpha$ between $0.00001$ and $0.999$ for the contours of the approximate likelihood surface, giving 15318 rows representative points (including 6 MLEs from the 6 fitted LGMs). 

Likelihoods for these representative parameters are evaluated by the  \fct{gpuLik::likfitLgmGpu} function below.  The values for \code{boxcox} are a sequence of 33 equally-spaced values between the 0.01th and 0.99th quantiles of the Wald CI, as well as the MLE $\hat\lambda$.  The \code{NparamPerIter} argument specifies that 400 likelihoods will be evaluated simultaneously, meaning a 40,000 by 100 matrix is needed to hold the 400 covariance matrices.  
The matrix $\text{ssqYX} = (y, X)^\T V^{-1} (y,X)$ returned by \fct{crossprodBatch} is of $36\cdot 400$ rows and $36$ columns,34 columns for the 34 Box Cox parameters and one column for each of the two covariates. 
\begin{knitrout}
\definecolor{shadecolor}{rgb}{1, 1, 1}\color{fgcolor}\begin{kframe}
\begin{verbatim}
R> result <- gpuLik::likfitLgmGpu(model = mle1, params = paramsUse,
+    paramToEstimate = 
+      c('range', 'combinedRange', 'shape', 'nugget', 'aniso1', 'aniso2', 'boxcox'),
+  	boxcox = seq(b[1], b[9], len = 33), cilevel = 0.9, 
+    NparamPerIter = 400, Nglobal = c(128, 128), Nlocal = c(16, 16), NlocalCache = 2800)
\end{verbatim}
\end{kframe}
\end{knitrout}

\begin{figure}[H]
\centering
	\subcaptionbox{$\sqrt{\phi_X \phi_Y}$ or $\gamma_1$\label{fig:m1}}{\includegraphics[width=0.44\textwidth]{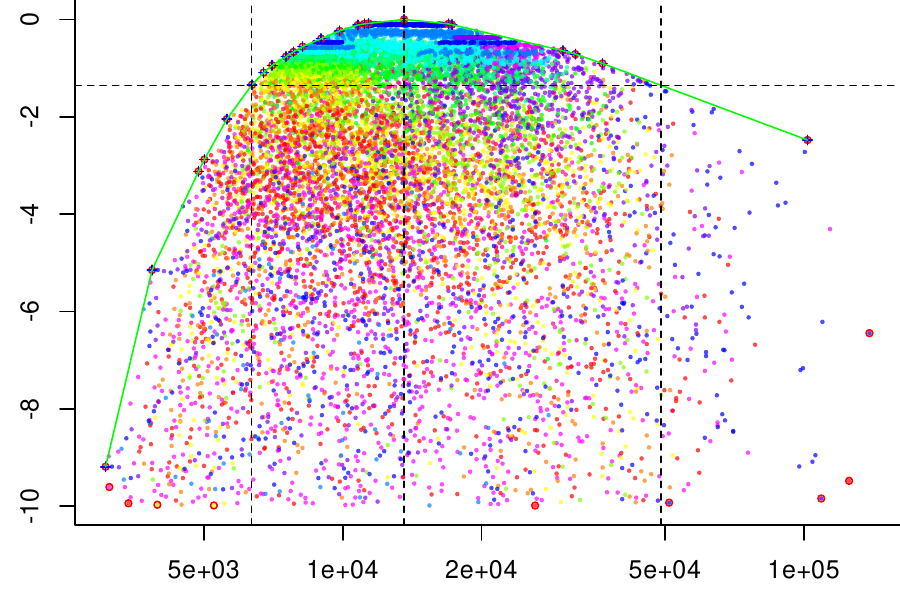} }
	\subcaptionbox{$\kappa$\label{fig:m2}}{\includegraphics[width=0.44\textwidth]{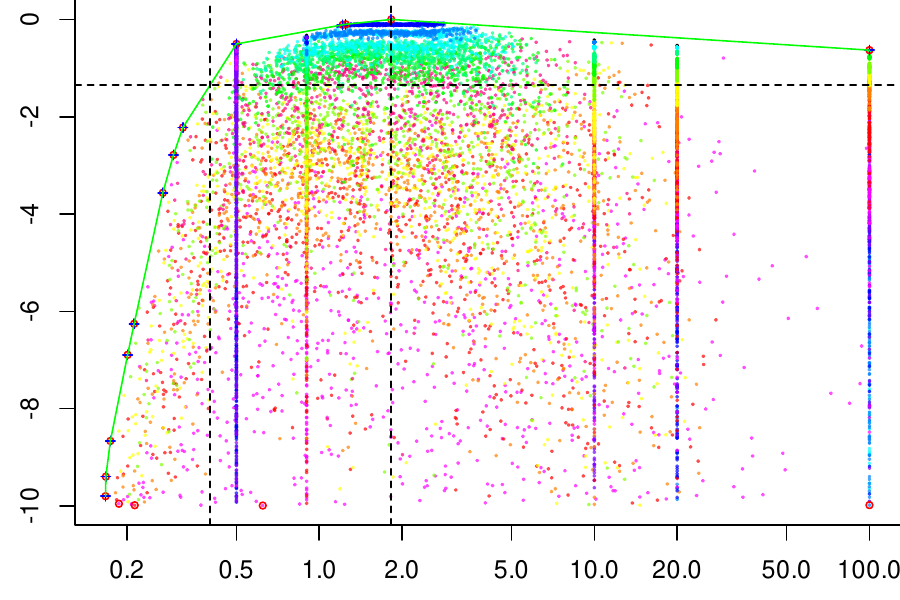} }
	\subcaptionbox{$\nu^2$\label{fig:m3}}{\includegraphics[width=0.44\textwidth]{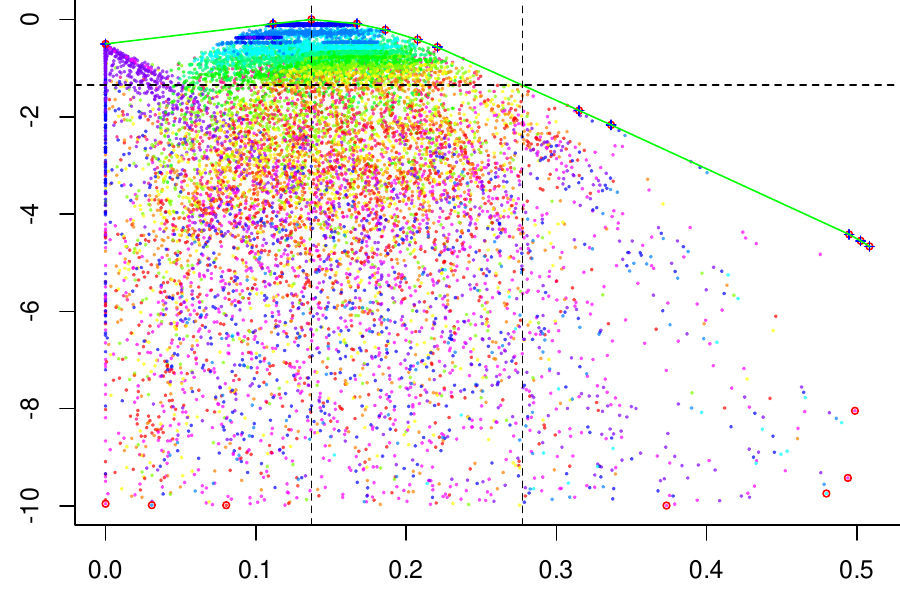} }
	\subcaptionbox{aniso1, $\gamma_2$\label{fig:m4}}{\includegraphics[width=0.44\textwidth]{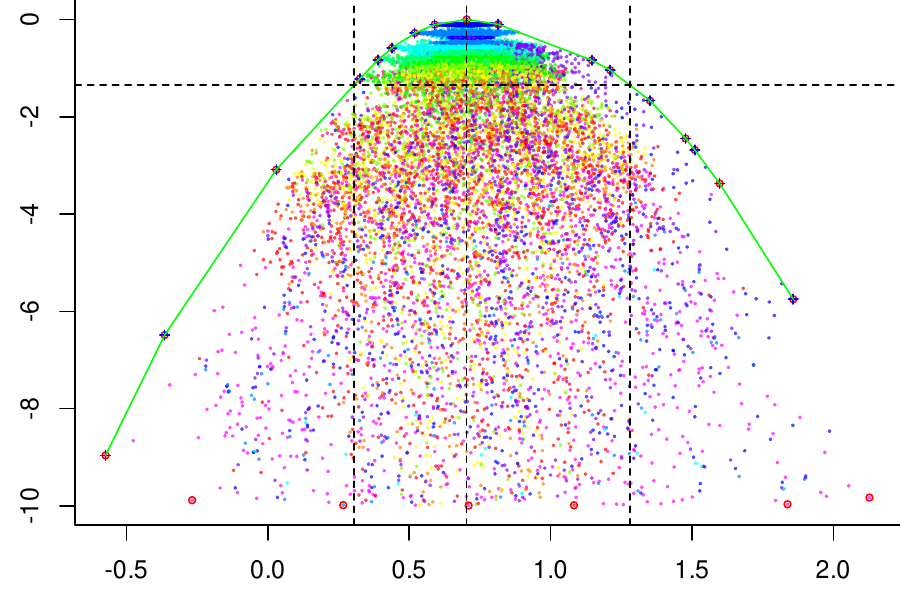} }
	\subcaptionbox{aniso2, $\gamma_3$\label{fig:m5}}{\includegraphics[width=0.44\textwidth]{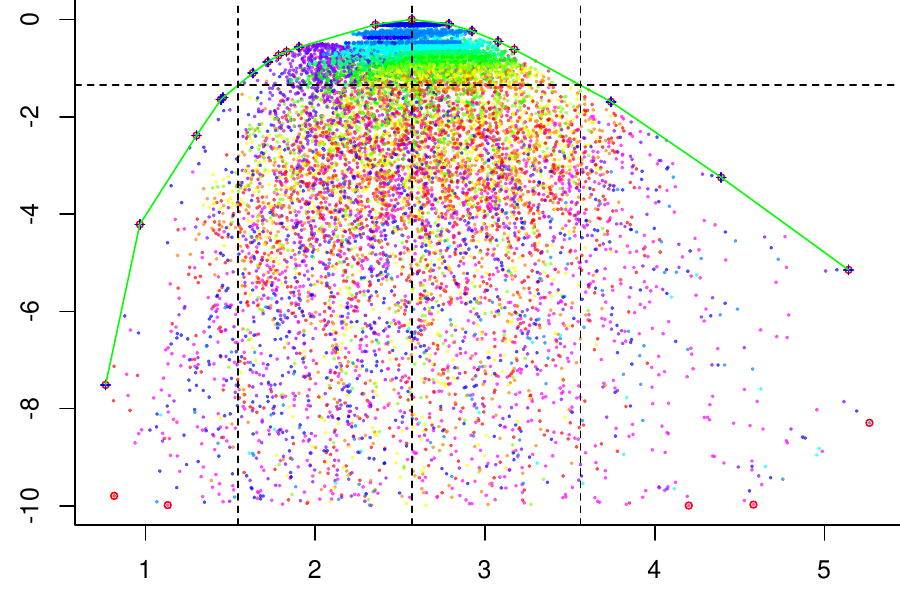} }
	\subcaptionbox{anisotropy ratio, $\phi_R$\label{fig:m6}}{\includegraphics[width=0.44\textwidth]{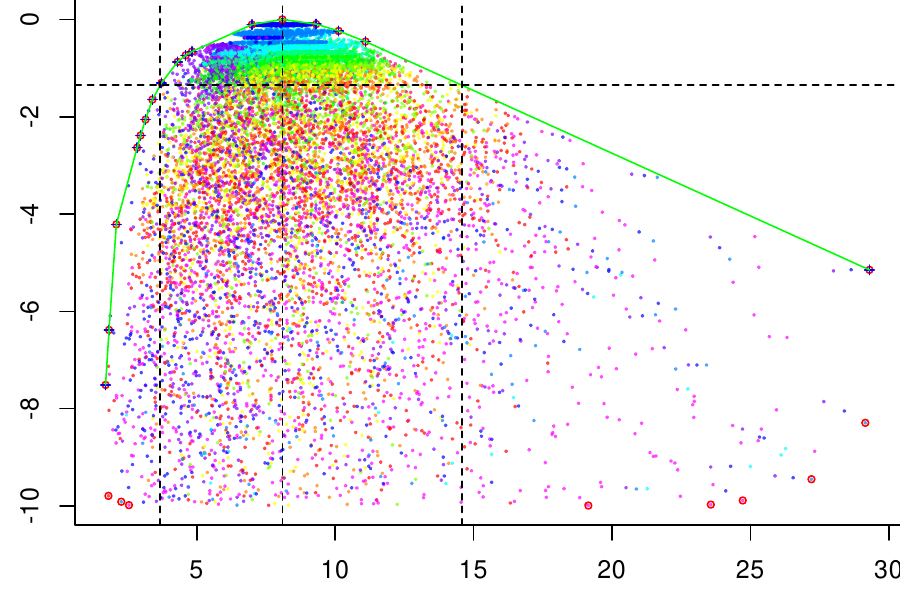} }
	\subcaptionbox{anisotropy radians, $\phi_A$\label{fig:m7}}{\includegraphics[width=0.44\textwidth]{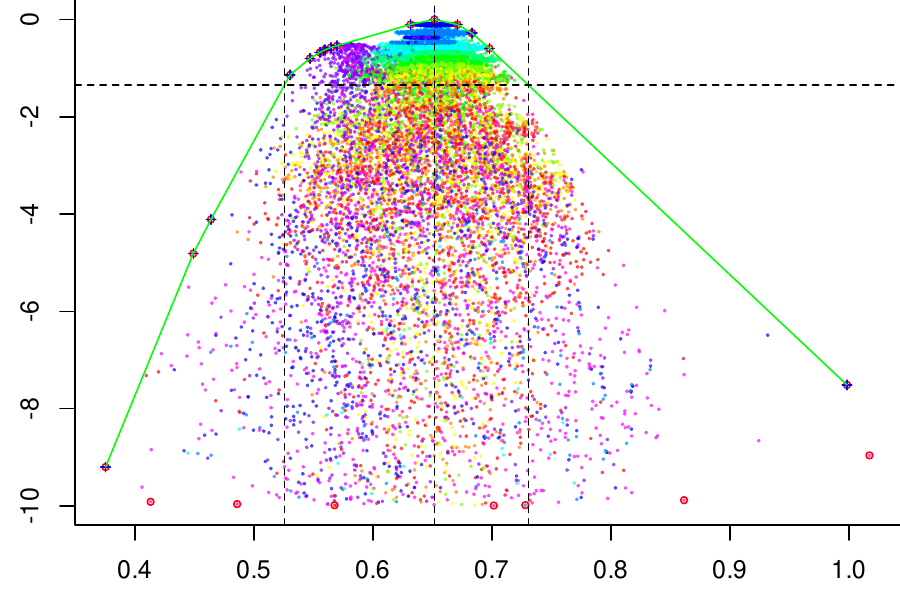} }
	\subcaptionbox{Legend\label{fig:m8}}{\includegraphics[width=0.44\textwidth]{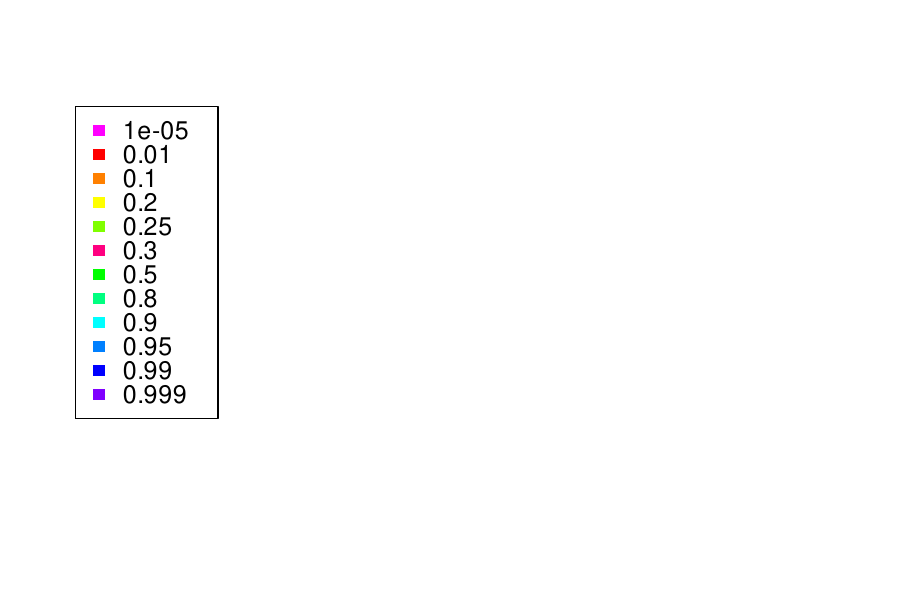} }
	\caption[]{Profile log-likelihoods for  the correlation parameters estimated from the Swiss rain data.  Vertical axes are log likelihood minus the likelihood at the MLE.  Dashed lines show the MLE and  90\% confidence intervals. Point clouds are representative points and their likelihoods, with colours showing the $\alpha$ quantile of the contour each point was drawn from.}\label{swissprofile1}
\end{figure}

The \code{NlocalCache} argument is related to the amount of local memory available on the GPU, and 2,800 real numbers will be stored in local memory at any one time.  Each work group consists of a grid of 16 by 16 work items, comprising an 8 by 4 grid of work groups (giving $16 \cdot 8$ by $16 \cdot 4$ work items in total).

Figure \ref{swissprofile1} contains profile likelihoods for covariance parameters.  The point clouds are the representative parameters and their log likelihoods (minus the log likelihood at the MLE), with colours corresponding to the significance level $\alpha$ of the contours the points were drawn from. 
The PLL curve for $\kappa$ displayed in Figure \ref{fig:m2} is quite flat on the right side of its maximum, and the quadratic approximation to the likelihood is poor.  The configuration points with $\kappa$ fixed at at 0.5, 0.9, 10, 20, 100 are clearly visible as vertical lines. The PLL for $\nu^2$ has a visible pile of points collected at 0 as we turn half of the negative $\nu^2$'s in the sample to be 0 and the other half are randomly sampled from the interval $(0,2)$.
The log-likelihoods for the reparametrized anisotropy parameters $\gamma_2$ and $\gamma_3$ are fairly close to quadratic, which supports the use of the polar coordinate transformation.  The data are clearly anisotropic although the ratio parameter has considerable uncertainty, possibly being as small as 4 or as large as 15.

 \begin{figure}[H]
		\centering
		\subcaptionbox{Box-Cox, $\lambda$}{\includegraphics[width=0.44\textwidth]{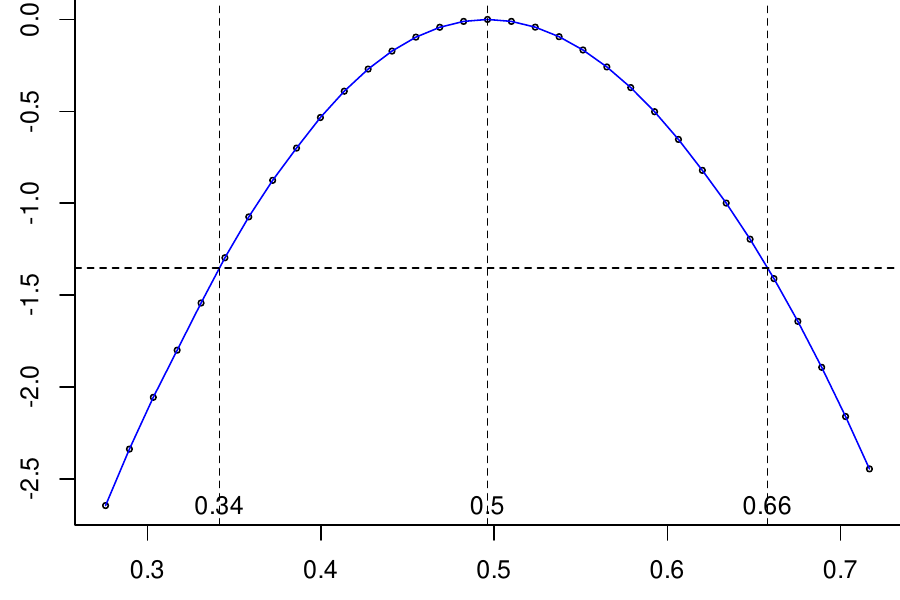}}
		\subcaptionbox{Standard deviation, $\sigma$}{\includegraphics[width=0.44\textwidth]{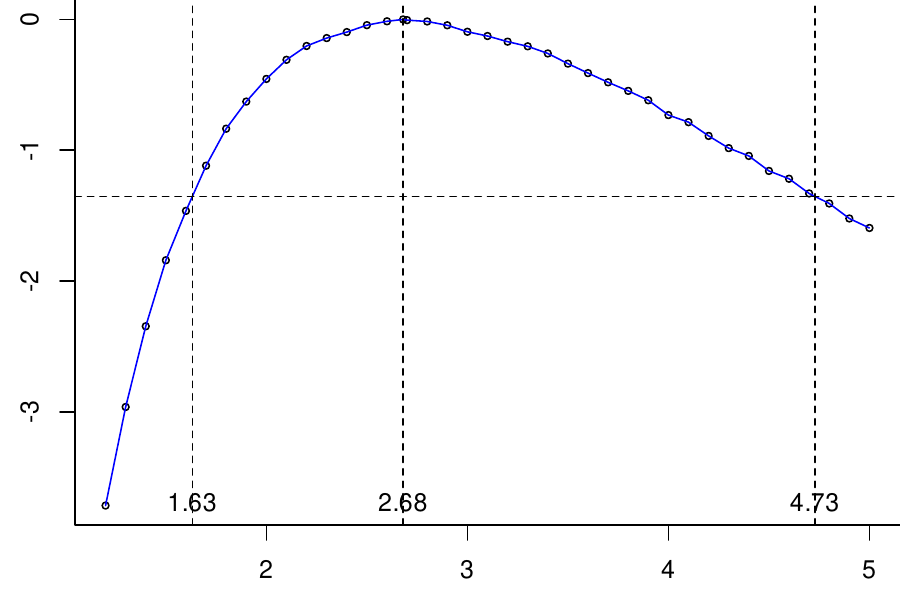}}
		\subcaptionbox{Intercept, $\beta_0$}{\includegraphics[width=0.44\textwidth]{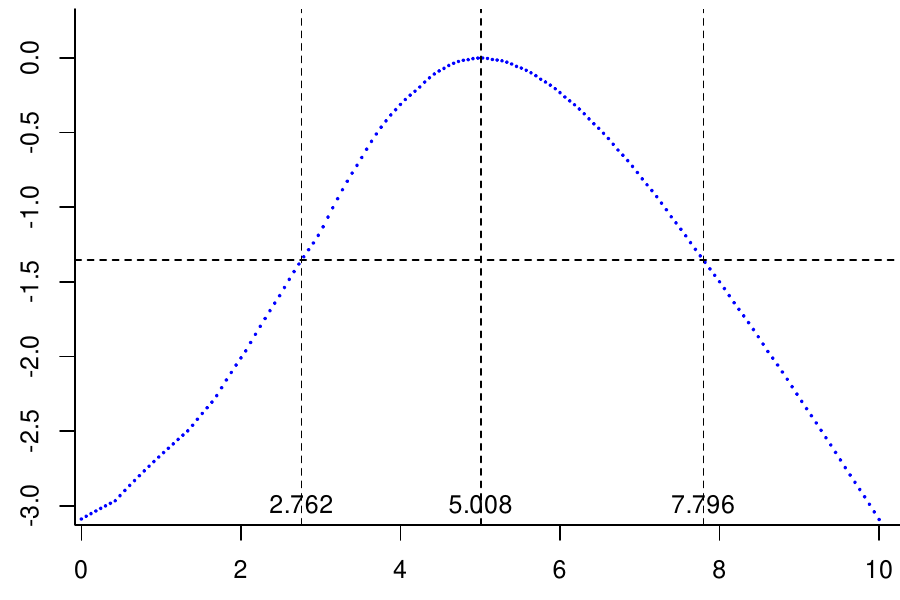}}
		\subcaptionbox{Elevation, $\beta_1 \cdot 1000$}{\includegraphics[width=0.44\textwidth]{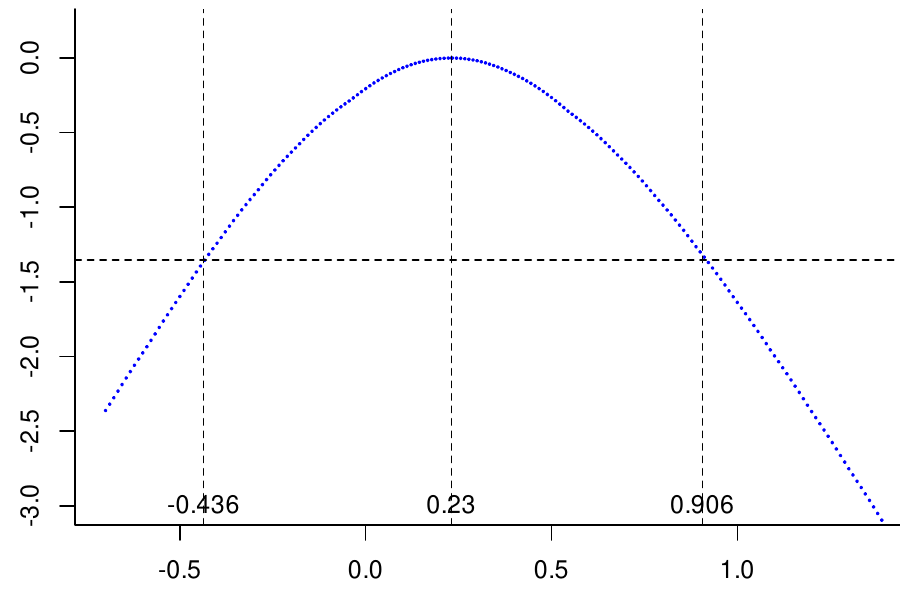}}
		\caption{Profile log-likelihoods for the Box Cox transformation parameter, standard deviation and regression parameters computed from the Swiss rain data. Horizontal dashed lines on each plot define approximate 90\% confidence intervals.}\label{swissprofile2}
	\end{figure}
	
\begin{figure}[tp]
	\centering
	\subcaptionbox{$(\beta_0, \beta_1 * 1000)$}{\includegraphics[width=0.44\textwidth]{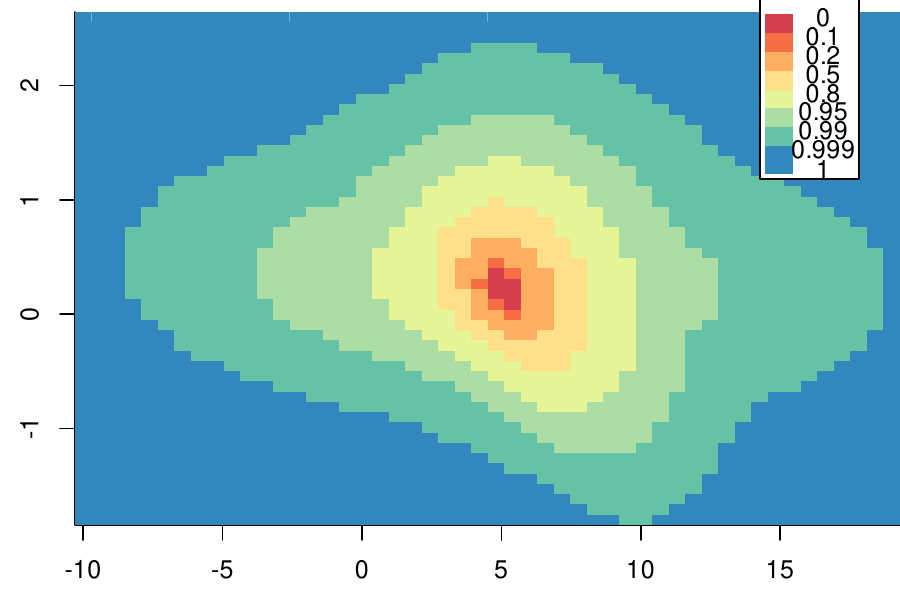}}
	\subcaptionbox{$(\phi_R, \phi_A)$}{\includegraphics[width=0.44\textwidth]{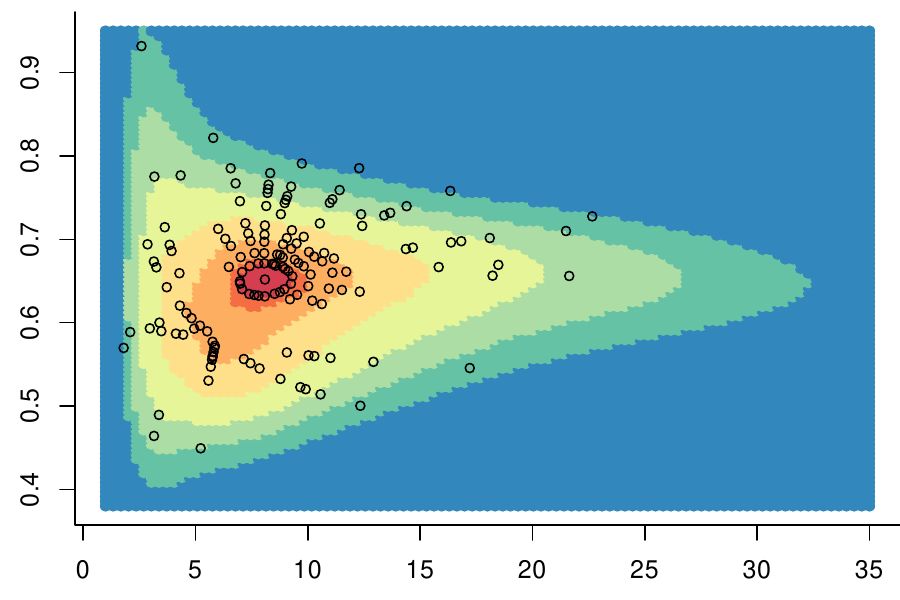}}
	\subcaptionbox{$(\nu^2, \gamma_2)$}{\includegraphics[width=0.44\textwidth]{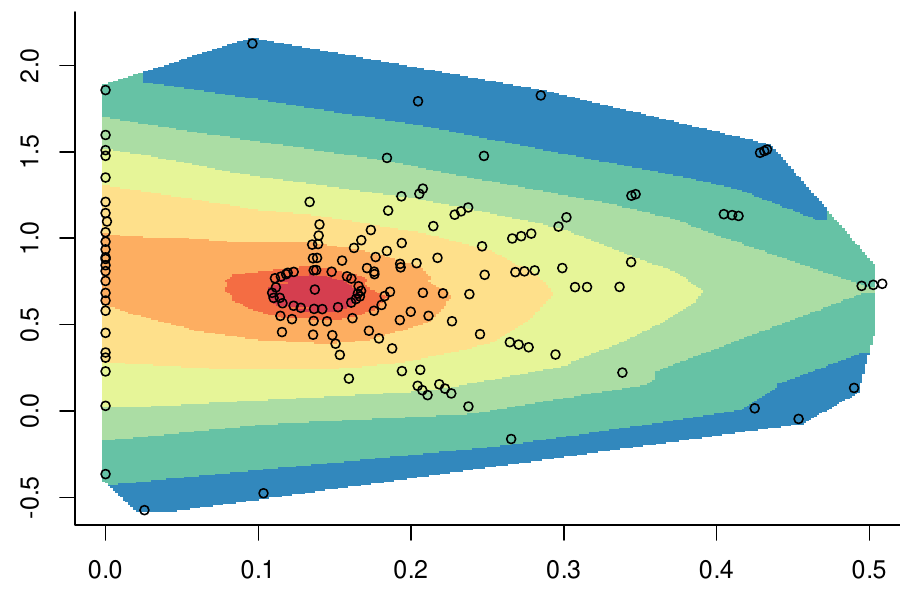}}
	\subcaptionbox{$(\gamma_1, \gamma_2)$}{\includegraphics[width=0.44\textwidth]{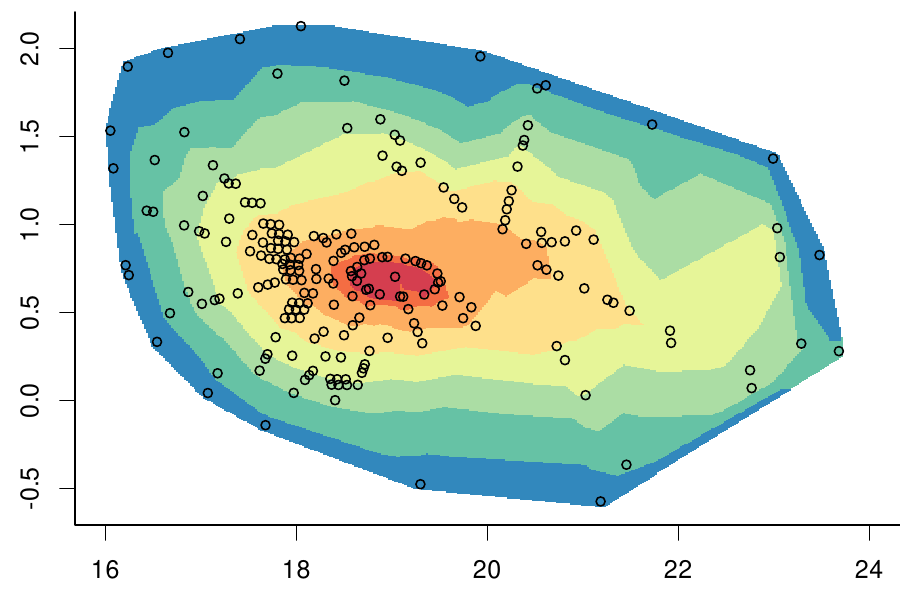}}
	\subcaptionbox{$(\gamma_2, \gamma_3)$}{\includegraphics[width=0.44\textwidth]{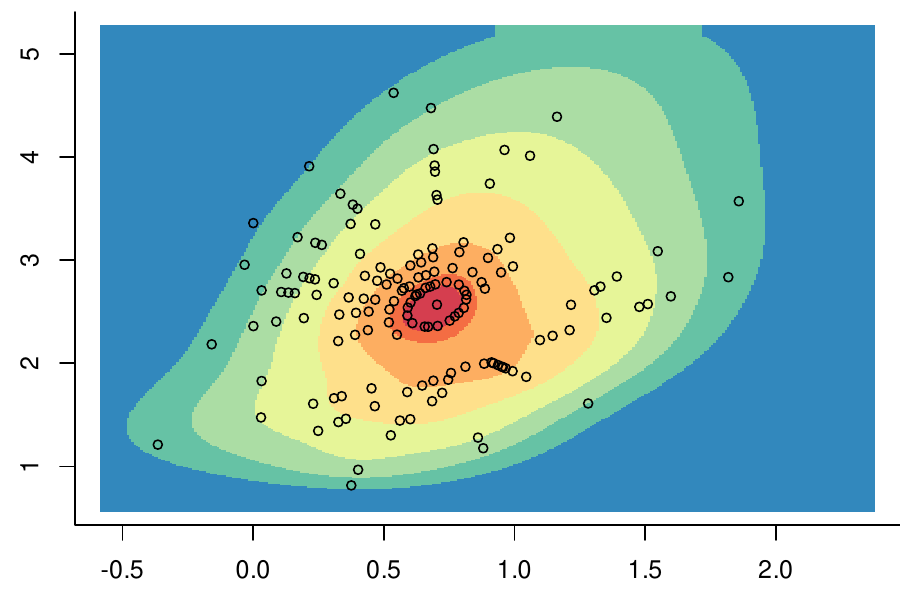}}
	\subcaptionbox{$(\gamma_1, \gamma_3)$}{\includegraphics[width=0.44\textwidth]{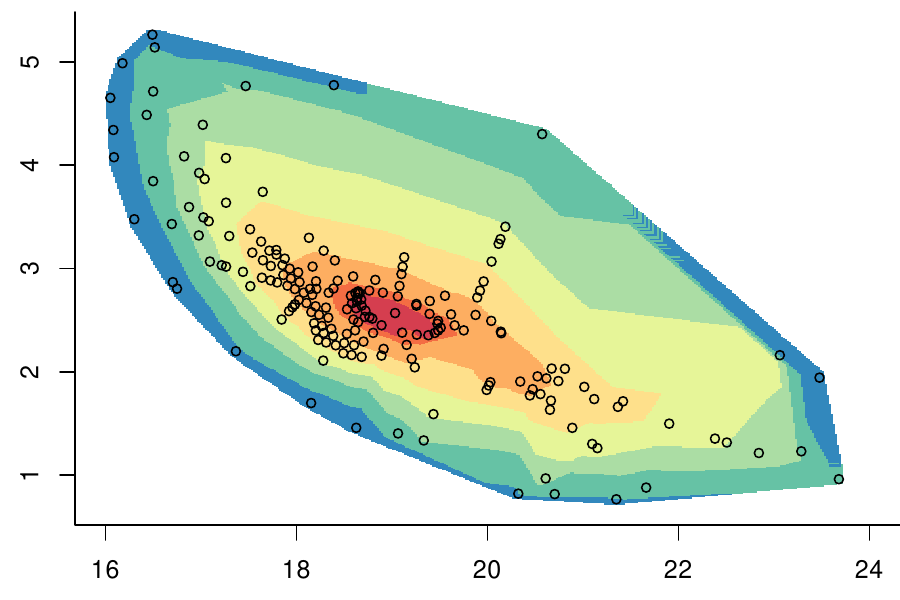}}
	\caption{Two-dimensional profile log-likelihood contour plots for  pairs of model parameters, computed from the Swiss rain data.  Circular plotting symbols show locations of representative points on convex hull.}\label{swissprofile3}
\end{figure}

The 2-dimensional PLL contour plots in  Figure \ref{swissprofile3}... of $(\gamma_2, \gamma_3)$ and $(\phi_R, \phi_A)$ in Figure \ref{swissprofile3} show the data are anisotropic. They demonstrate the advantages of the reparameterization, the contours of the internal parameters has a more oval-like shape compared to the PLL contours of the natural parameters. As the value of anisotropy ratio gets closer to 1 (isotropic), the estimate of CI for the anisotropy radians would become wider and eventually on the whole interval of $(-\pi/2, \pi/2)$. 

Table \ref{tab:swisssummary} shows the profile-based CIs from \pkg{gpuLik} are wider than Wald based CIs from \pkg{geostatsp}.
The CIs for $\lambda$ obtained by the two methods are very close.

Likelihood-based confidence intervals for model parameters are compared with \pkg{geostatsp}'s Wald-based CI's  in Table \ref{tab:swisssummary}. PLL plots for the model parameters and some 2-dimensional PLL plots are displayed in \Cref{swissprofile1,swissprofile2,swissprofile3}. 

\begin{knitrout}
\definecolor{shadecolor}{rgb}{1, 1, 1}\color{fgcolor}\begin{table}[H]

\caption{\label{tab:swisssummary}Swiss rainfall data. Comparison between Wald and profile likelihood based CI's.\label{swissTable}}
\centering
\begin{tabular}[t]{lcccccc}
\toprule
\multicolumn{3}{c}{ } & \multicolumn{2}{c}{likelihood-based} & \multicolumn{2}{c}{Wald} \\
\cmidrule(l{3pt}r{3pt}){4-5} \cmidrule(l{3pt}r{3pt}){6-7}
  & notation & estimate & ci0.05 & ci0.95 & ci0.05 & ci0.95\\
\midrule
(Intercept) & $\beta_1$ & 5.01 & 2.76 & 7.8 & 3.32 & 6.69\\
elevation*1000 & $\beta_2*1000$ & 0.23 & -0.44 & 0.91 & -0.36 & 0.82\\
sdSpatial & $\sigma$ & 2.68 & 1.63 & 4.74 & 1.57 & 4.58\\
range/1000 & $\phi_X/1000$ & 38.62 & 21.79 & 118.63 & 18.54 & 80.42\\
combinedRange/1000 & $\sqrt{\phi_X \phi_Y}/1000$ & 13.58 & 6.34 & 48.96 & NA & NA\\
anisoRatio & $\phi_R$ & 8.09 & 3.68 & 14.58 & 4.73 & 13.86\\
shape & $\kappa$ & 1.83 & 0.40 & $>100$ & -1.25 & 4.91\\
nugget & $\nu^2$ & 0.14 & 0.00 & 0.28 & NA & NA\\
sdNugget & $\tau$ & 0.99 & 0.00 & 1.41 & 0.62 & 1.60\\
anisoAngleRadians & $\phi_{A}$ & 0.65 & 0.53 & 0.73 & 0.58 & 0.73\\
aniso1 & $\gamma_2$ & 0.70 & 0.30 & 1.28 & NA & NA\\
aniso2 & $\gamma_3$ & 2.57 & 1.55 & 3.56 & NA & NA\\
boxcox & $\lambda$ & 0.50 & 0.34 & 0.66 & 0.34 & 0.65\\
\bottomrule
\end{tabular}
\end{table}

\end{knitrout}


\section{Example: Soil mercury data} \label{soil}

\begin{figure}[H]
	\centering
	\subcaptionbox{Mercury in soil\label{fig:mapsoil}}{\includegraphics[width=0.33\textwidth]{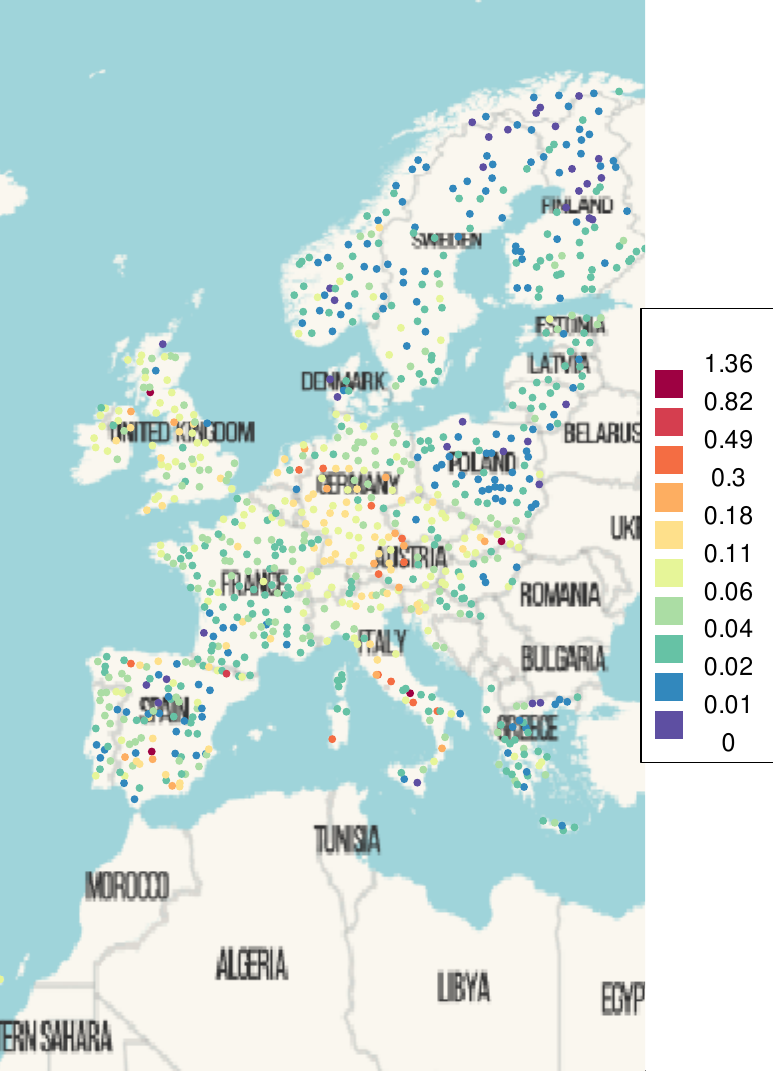}}
	\subcaptionbox{Land type\label{fig:mapsoilB}}{\includegraphics[width=0.33\textwidth]{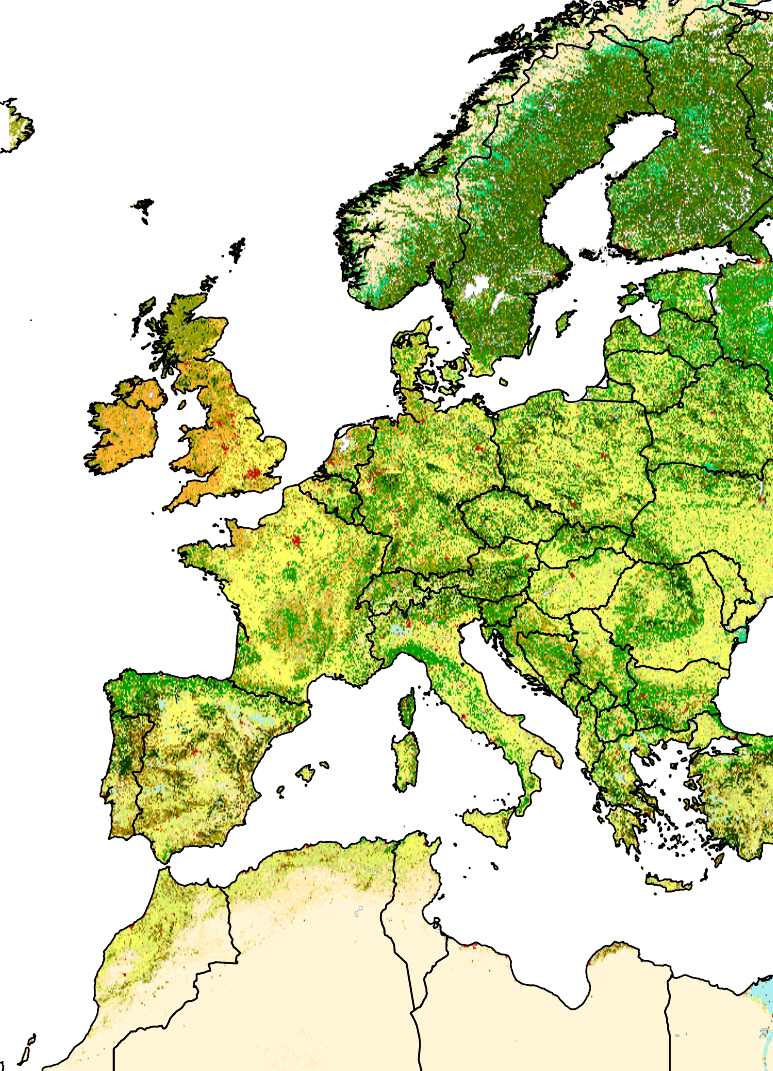}}
	\subcaptionbox{Land type\label{fig:mapsoilC}}{\includegraphics[width=0.32\textwidth]{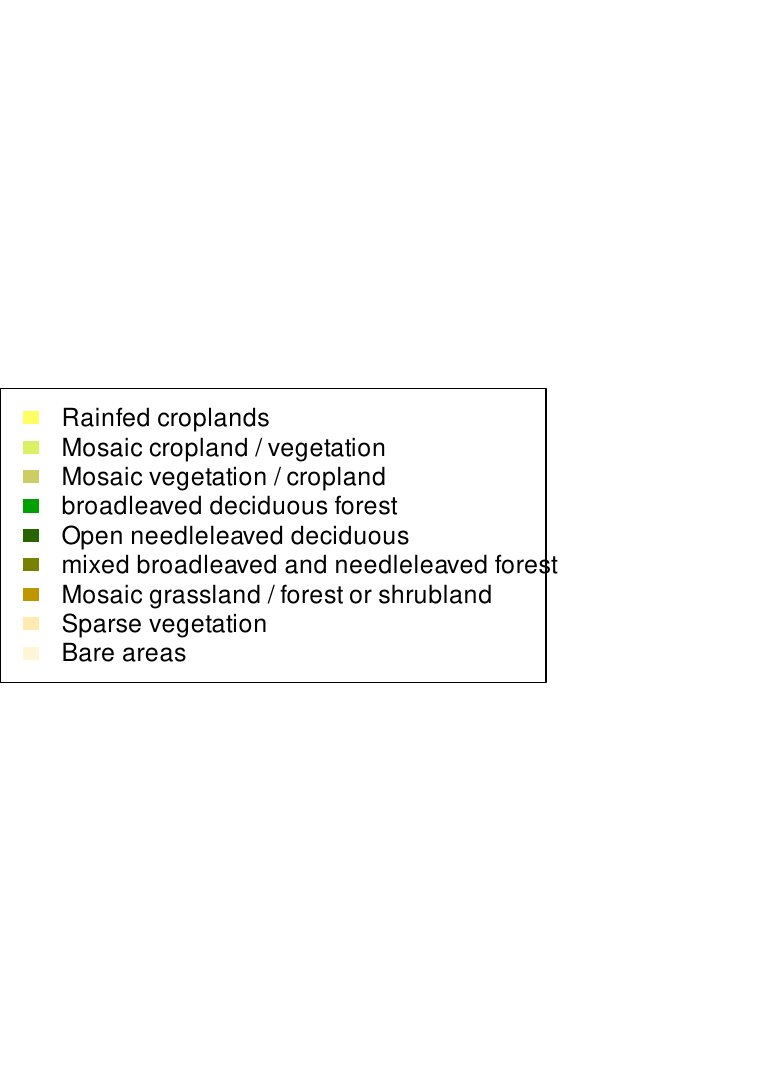}}
	\caption{Soil mercury measurement with locations in Europe.}\label{soilmaps}
\end{figure}

To illustrate our method with a larger data set, we apply the method on the considerably larger soil mercury dataset which has 829 observations and 29 predictors. The soil mercury data is 
obtained from ``The EuroGeoSurveys - FOREGS Geochemical Baseline Database" \cite{salminen2005geochemical}, available for download at \url{http://weppi.gtk.fi/publ/foregsatlas/ForegsData.php}. The mercury concentration measurement in topsoils is plotted in Figure \ref{fig:mapsoil} 
The covariate $X$ contains: elevation, land category (see Figure \ref{fig:mapsoilB} and \ref{fig:mapsoilC}), night-time lights, vegetation index. 

The procedure is identical to that used in the Swiss rainfall data example. We fit 4 models 
using \fct{geostatsp::lgm} to obtain the MLEs, the first using the MLE of the Mat\'ern shape parameter and final three fixing the parameter at 0.5, 0.8, and 1. \fct{gpuLik::configParams} configured the candidate points  based on the MLEs of each model. The contour levels selected are the same as in the Swiss rainfall data example, the parameter configuration matrix has 12316 rows (including MLEs from the 4 fitted LGMs, $12\cdot 726$ configurations based on the \textit{hgRes1} object below, and $3\cdot 10 \cdot 120$ configurations based on the other 3 models) and 5 columns. The values for \code{boxcox} is configured by choosing 31 equally-spaced values between the 0.01th (\code{b[1]}) and 0.99th (\code{b[2]}) quantiles of the approximated Gaussian distribution. We obtained the log-likelihoods by \fct{gpuLik::likfitLgmGpu} with the code below.

\begin{knitrout}
\definecolor{shadecolor}{rgb}{1, 1, 1}\color{fgcolor}\begin{kframe}
\begin{verbatim}
R> hgRes1 = geostatsp::lgm(formula = HG ~ elevation + land + night + evi,
+      data = hgm, grid = 20, covariates = covList, fixBoxcox = FALSE,
+      fixShape = FALSE, fixNugget = FALSE, reml = FALSE, aniso = TRUE)
\end{verbatim}
\end{kframe}
\end{knitrout}

\begin{knitrout}
\definecolor{shadecolor}{rgb}{1, 1, 1}\color{fgcolor}\begin{kframe}
\begin{verbatim}
R> reParamaters <- gpuLik::configParams(list(hgRes1, hgRes2, hgRes3, hgRes4),
+      alpha = c(0.00001, 0.01, 0.1, 0.2, 0.25, 0.3, 0.5, 0.8, 0.9, 0.95,
+          0.99, 0.999))
R> paramsUse <- reParamaters$representativeParamaters[, 1:5]
R> b <- reParamaters$boxcox
\end{verbatim}
\end{kframe}
\end{knitrout}

\begin{knitrout}
\definecolor{shadecolor}{rgb}{1, 1, 1}\color{fgcolor}\begin{kframe}
\begin{verbatim}
R> result <- gpuLik::likfitLgmGpu(model = hgRes1, params = paramsUse,
+      boxcox = seq(b[1], b[9], len = 31), cilevel = 0.95, type = "double",
+      convexHullForBetas = FALSE, NparamPerIter = 400, Nglobal = c(256,
+          256), Nlocal = c(16, 16), NlocalCache = 2800, verbose = c(1,
+          0))
\end{verbatim}
\end{kframe}
\end{knitrout}

\begin{figure}[tp]
	\centering
	\subcaptionbox{$\sqrt{\phi_X \phi_Y}$, $\gamma_1$}{\includegraphics[width=0.44\textwidth]{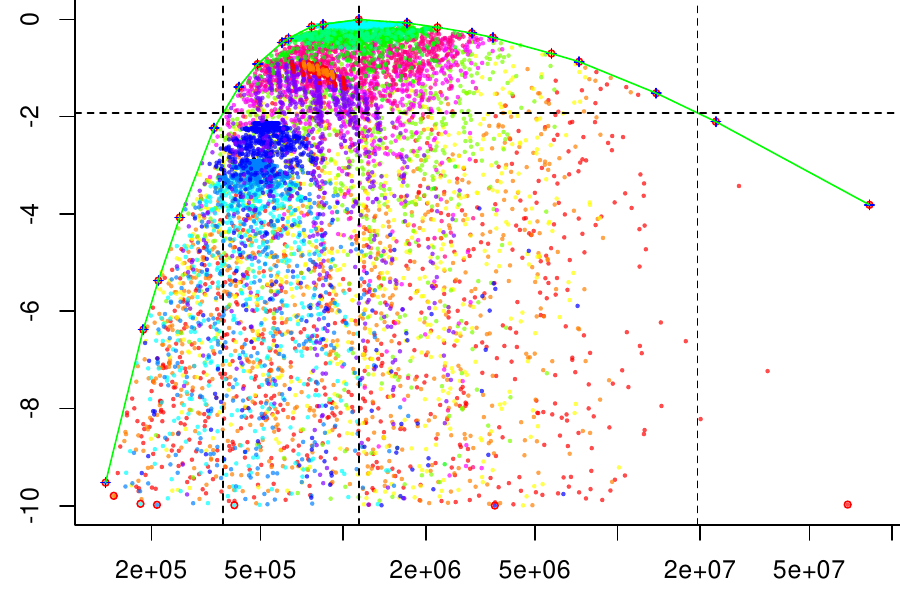}}
	\subcaptionbox{$\kappa$}{\includegraphics[width=0.44\textwidth]{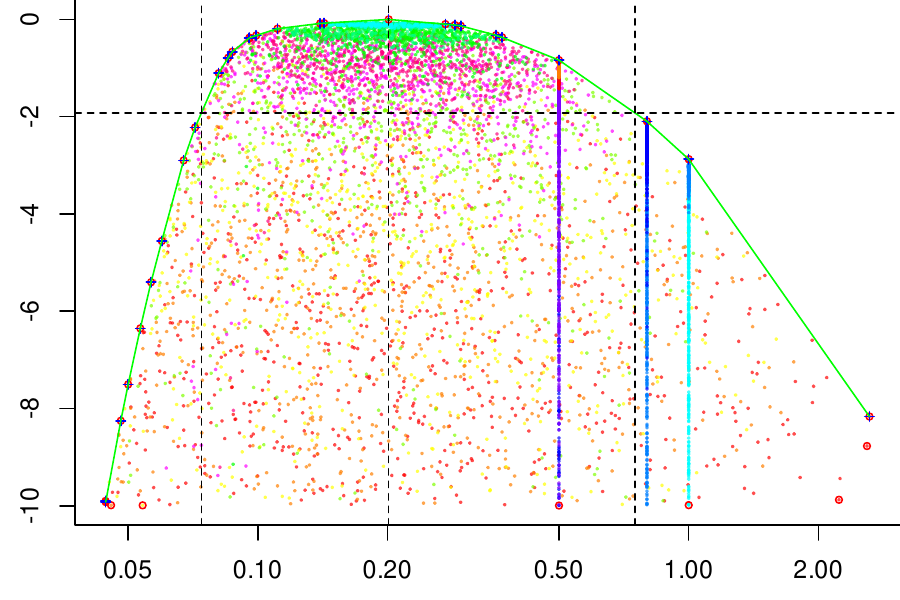}}
	\subcaptionbox{$\nu^2$}{\includegraphics[width=0.44\textwidth]{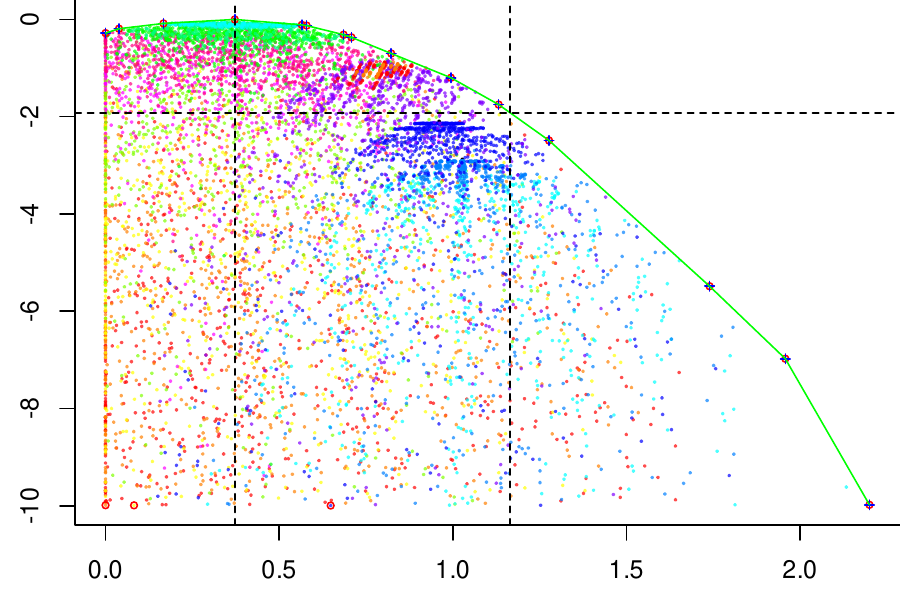}}
	\subcaptionbox{$\gamma_2$}{\includegraphics[width=0.44\textwidth]{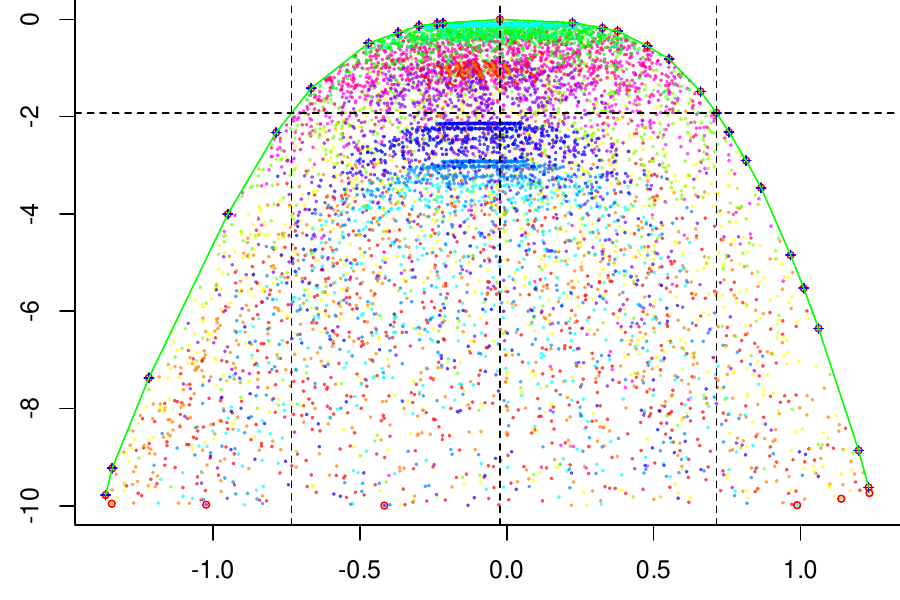}}
	\subcaptionbox{$\gamma_3$}{\includegraphics[width=0.44\textwidth]{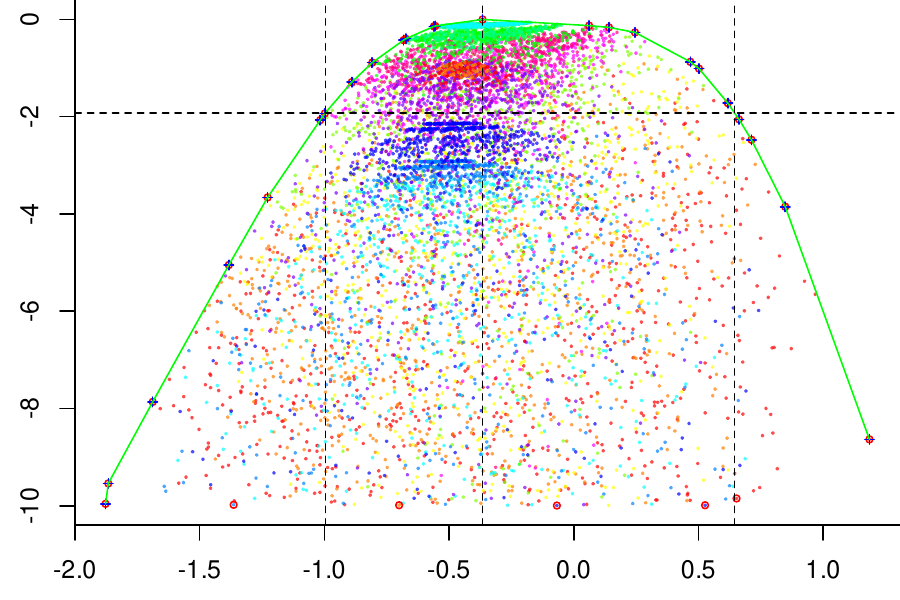}}
	\subcaptionbox{anisotropy ratio, $\phi_R$}{\includegraphics[width=0.44\textwidth]{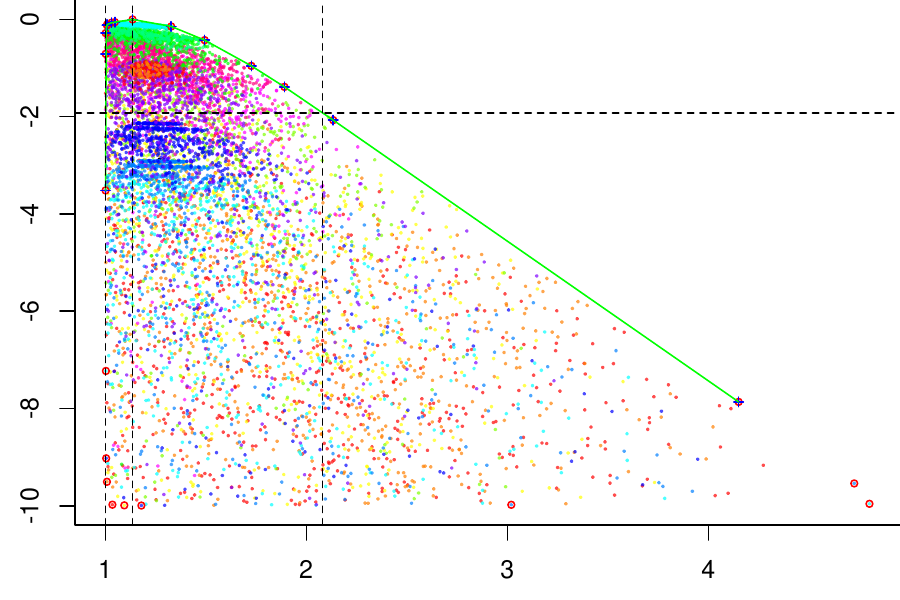}}
	\subcaptionbox{anisotropy radians, $\phi_A$}{\includegraphics[width=0.44\textwidth]{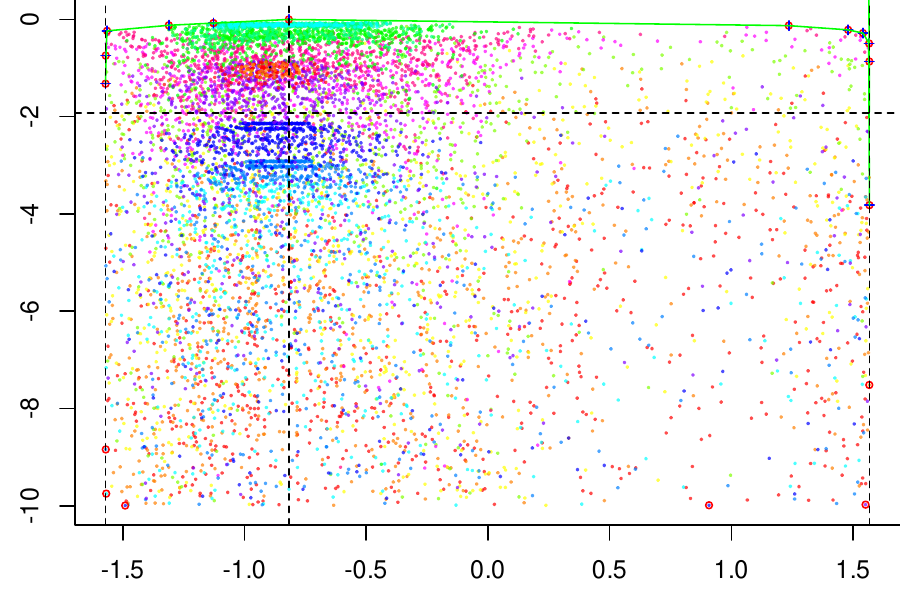}}
	\subcaptionbox{Legend}{\includegraphics[width=0.44\textwidth]{swiss/legend.pdf}}
	\caption{Profile log-likelihoods for  the correlation parameters estimated from the Soil mercury data.  Vertical axes are log likelihood minus the likelihood at the MLE.  Dashed lines show the MLE and  95\% confidence intervals. Point clouds are representative points and their likelihoods, with colours showing the $\alpha$ quantile of the contour each point was drawn from.
	}\label{hgprofile1}
\end{figure} 

\begin{figure}[tp]
	\centering 
	\subcaptionbox{$(\gamma_2,\gamma_3)$}{\includegraphics[width=0.44\textwidth]{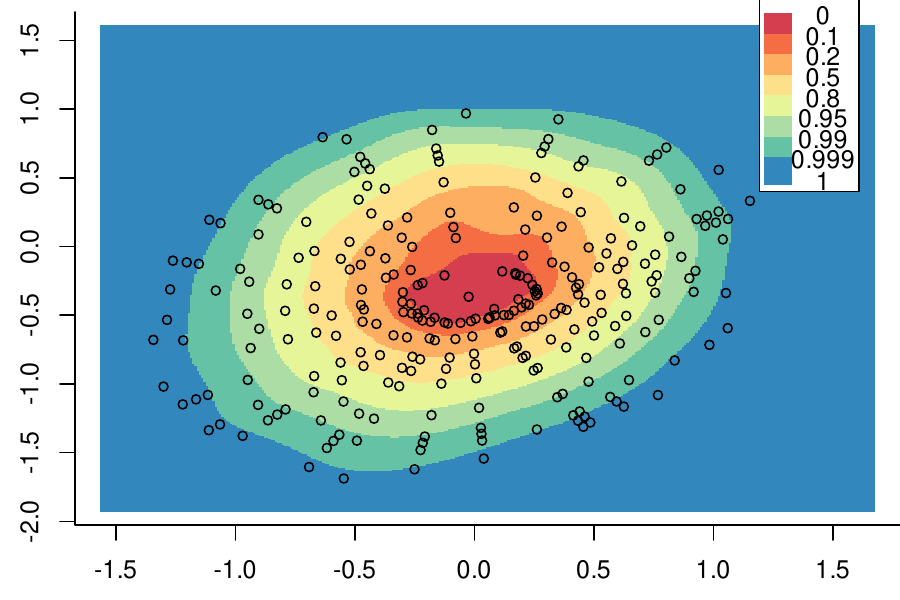}}
	\subcaptionbox{$(\phi_R,\phi_A)$}{\includegraphics[width=0.44\textwidth]{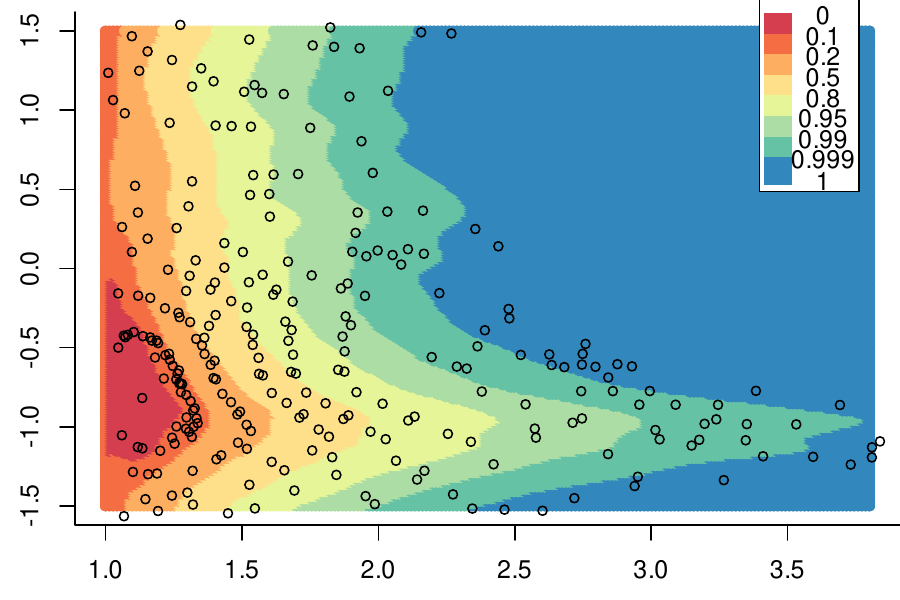}}
	\subcaptionbox{$(\gamma_1,\nu^2)$}{\includegraphics[width=0.44\textwidth]{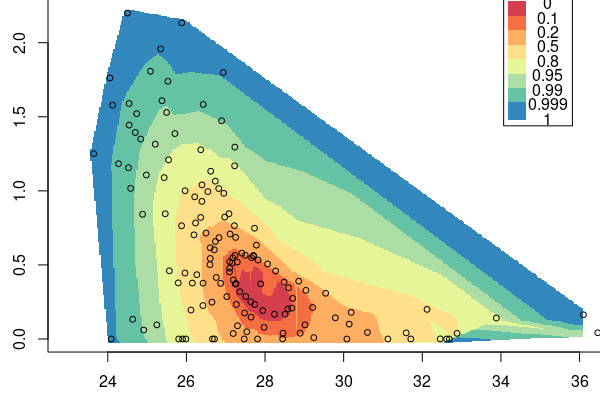}}
	\subcaptionbox{$(\gamma_1,\gamma_2)$}{\includegraphics[width=0.44\textwidth]{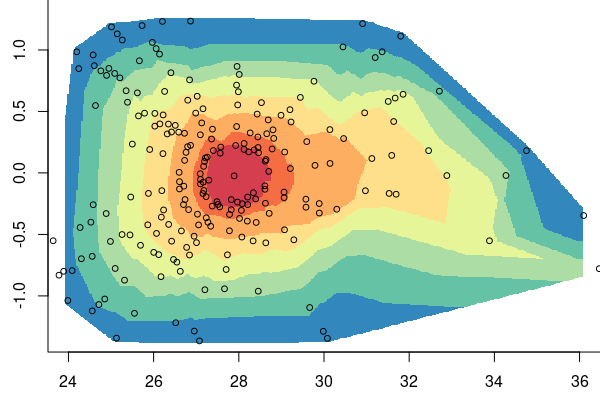}}
	\subcaptionbox{$(\gamma_1,\gamma_3)$}{\includegraphics[width=0.44\textwidth]{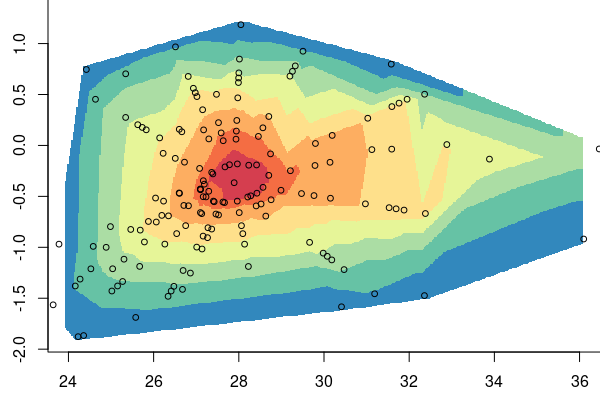}} 
	\subcaptionbox{$(\gamma_1,\log(\kappa))$}{\includegraphics[width=0.44\textwidth]{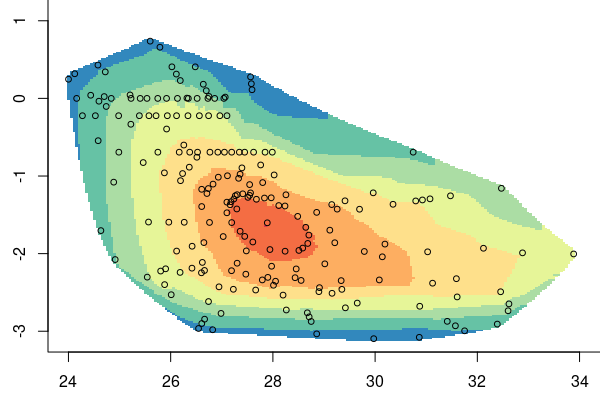}}  
	\subcaptionbox{$(\nu^2,\gamma_2)$}{\includegraphics[width=0.44\textwidth]{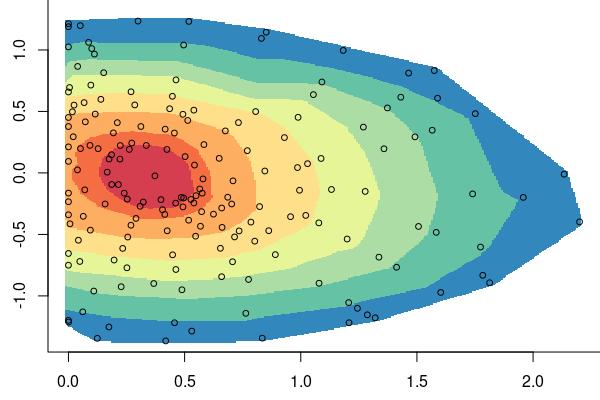}}  
	\subcaptionbox{$(\nu^2,\gamma_3)$}{\includegraphics[width=0.44\textwidth]{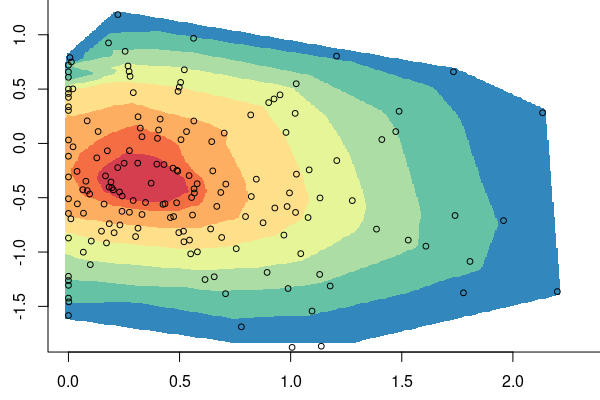}}  
	\caption{Two-dimensional profile log-likelihood contour plots for pairs of model parameters, computed from the Soil mercury data. Circular plotting symbols show locations of representative points on convex hull.
	}\label{hgprofile2}
\end{figure}

The soil mercury data is near isotropic, and the CI for the anisotropy radians parameter covers the entire domain $(-\pi/2, \pi/2)$. 
The 2-dimensional profile log-likelihood contour plots (a), (b) in Figure \ref{hgprofile2}, using the internal parameters does make the quadratic approximation for the likelihoods much better. Parameter estimates and CIs are compared to the Wald CI's in Table \ref{tab:soilcompare2}. For most of the parameters, profile likelihood gives wider CIs than Wald does, for parameters such as anisotropy radians, anisotropy ratio and shape, \pkg{gpuLik} gives better CIs as it takes into account the parameter boundaries. Again, very similar CI's for the transformation parameter are obtained by the two methods.
 
\begin{figure}[tp]
	\centering
	\subcaptionbox{Standard deviation, $\sigma$}{\includegraphics[width=0.44\textwidth]{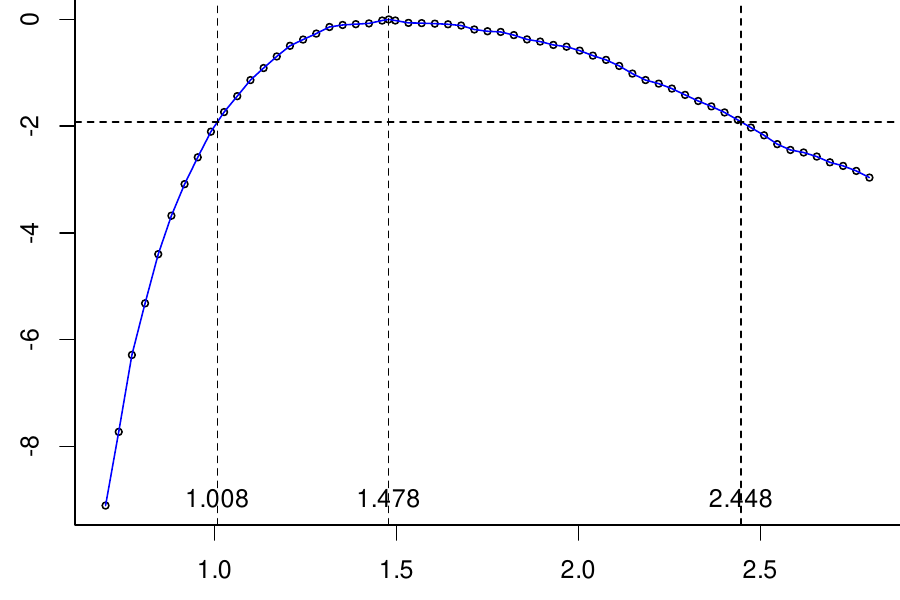}}
	\subcaptionbox{Intercept, ($\beta_0$)}{\includegraphics[width=0.44\textwidth]{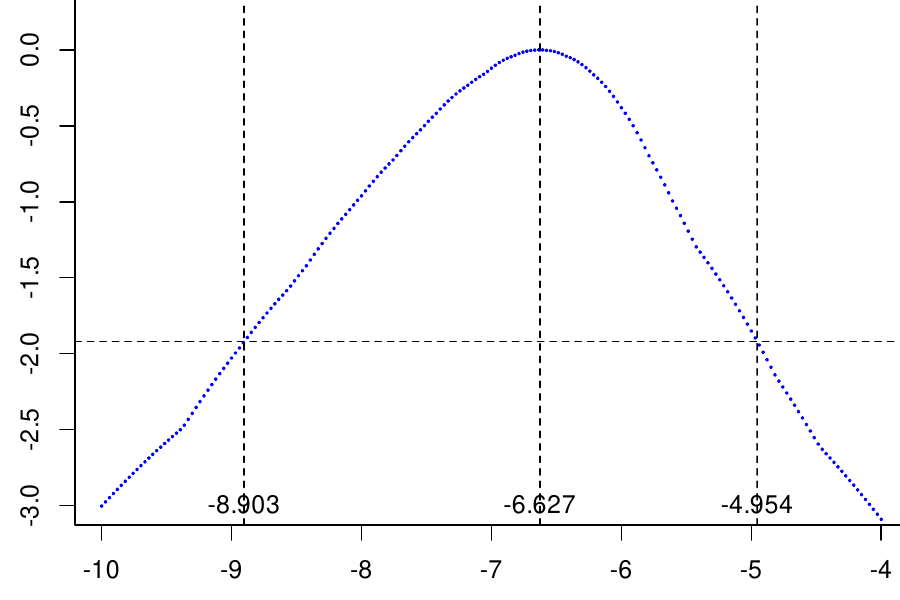}}
	\subcaptionbox{Elevation,$\beta_1 *10000$}{\includegraphics[width=0.44\textwidth]{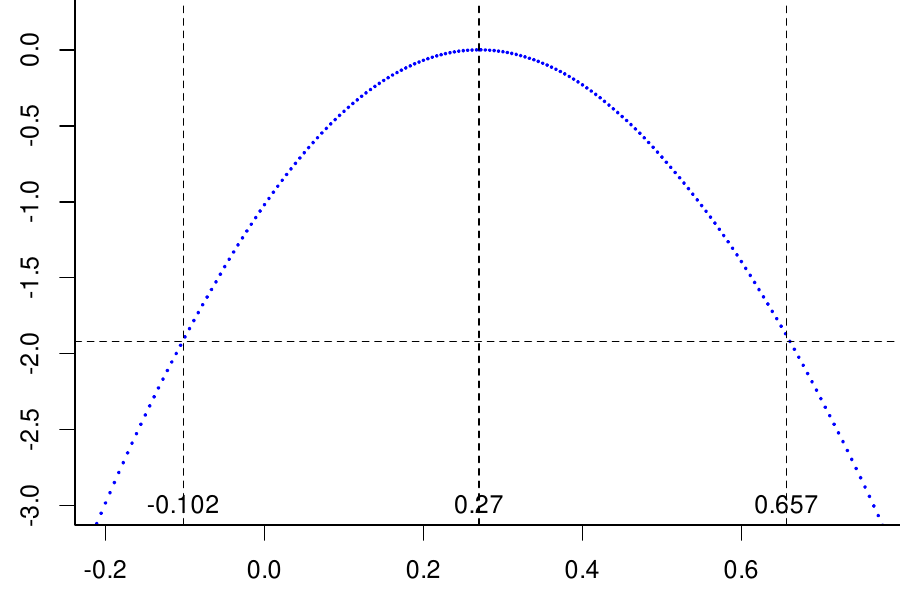}}
	\subcaptionbox{Box-Cox, $\lambda$}{\includegraphics[width=0.44\textwidth]{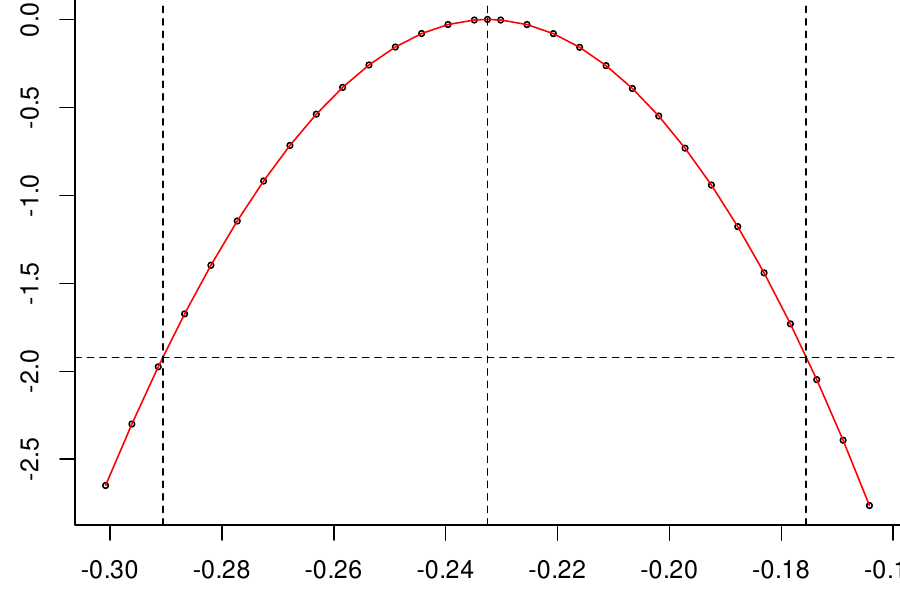}}
	\subcaptionbox{night, $\beta_{16}$}{\includegraphics[width=0.44\textwidth]{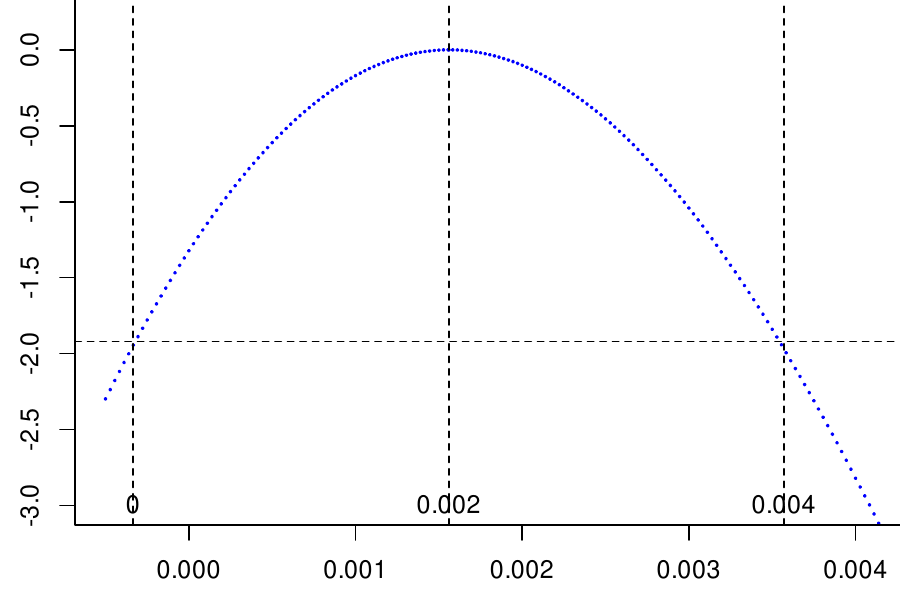}}
	\subcaptionbox{landMosaic\_grassland,$\beta_{5}$}{\includegraphics[width=0.44\textwidth]{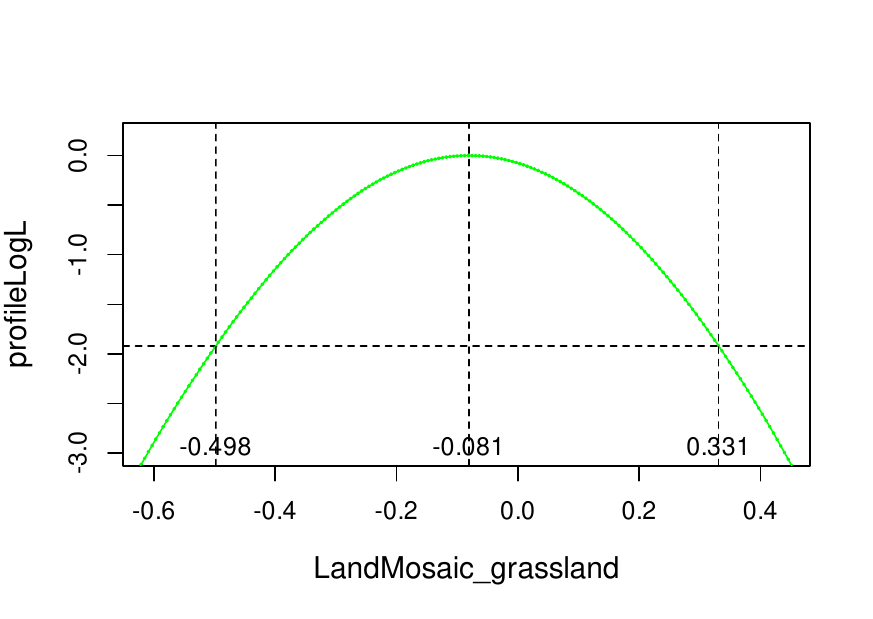}}
	\caption{Profile log-likelihoods for the standard deviation, some of the regression parameters, and Box Cox transformation parameter computed from the Soil mercury data. Horizontal dashed lines on each plot define approximate 95\% confidence intervals and the MLE.
		}\label{hgprofileBeta}
\end{figure}

\begin{knitrout}
\definecolor{shadecolor}{rgb}{1, 1, 1}\color{fgcolor}\begin{table}[H]

\caption{\label{tab:soilcompare2}95 \% CI's for model parameters in the LGM fit to the Soil mercury data. Shown are the GPU-computed profile likelihood CI's from \pkg{gpuLik} and the traditional Wald-based CI's computed by \pkg{geostatsp}.}
\centering
\begin{tabular}[t]{>{\raggedright\arraybackslash}p{9em}cccccc}
\toprule
\multicolumn{3}{c}{ } & \multicolumn{2}{c}{likelihood-based} & \multicolumn{2}{c}{Wald} \\
\cmidrule(l{3pt}r{3pt}){4-5} \cmidrule(l{3pt}r{3pt}){6-7}
  & notation & estimate & ci0.025 & ci0.975 & ci0.025 & ci0.975\\
\midrule
(Intercept) & $\beta_1$ & -6.63 & -8.90 & -4.98 & -7.80 & -5.45\\
elevation*10000 & $\beta_2*10000$ & 2.72 & -1.02 & 6.57 & -1.00 & 6.40\\
broadleaved deciduous forest & $\beta_3$ & -0.28 & -0.65 & 0.08 & -0.63 & 0.08\\
Mosaic cropland / vegetation & $\beta_4$ & -0.08 & -0.43 & 0.25 & -0.42 & 0.25\\
Open needleleaved deciduous & $\beta_5$ & -0.63 & -1.21 & -0.10 & -1.16 & -0.10\\
Mosaic grassland & $\beta_6$ & -0.08 & -0.50 & 0.33 & -0.49 & 0.33\\
Mosaic vegetation / cropland & $\beta_7$ & -0.37 & -0.80 & 0.03 & -0.78 & 0.03\\
mixed broadleaved & $\beta_8$ & -0.51 & -1.01 & -0.04 & -0.97 & -0.05\\
Sparse vegetation & $\beta_9$ & -1.02 & -1.60 & -0.50 & -1.52 & -0.52\\
herbaceous vegetation & $\beta_{10}$ & 0.00 & -0.58 & 0.59 & -0.57 & 0.58\\
Mosaic forest or shrubland & $\beta_{11}$ & -0.28 & -0.91 & 0.32 & -0.88 & 0.31\\
needleleaved evergreen forest & $\beta_{12}$ & -0.01 & -0.60 & 0.57 & -0.58 & 0.56\\
shrubland & $\beta_{13}$ & 0.20 & -0.49 & 0.90 & -0.48 & 0.88\\
grassland or woody vegetation & $\beta_{14}$ & -0.69 & -1.64 & 0.20 & -1.58 & 0.20\\
Artificial surfaces & $\beta_{15}$ & -0.44 & -2.38 & 1.47 & -2.32 & 1.45\\
Water bodies & $\beta_{16}$ & -1.17 & -3.78 & 1.35 & -3.67 & 1.34\\
night*1000 & $\beta_{17}*1000$ & 1.57 & -0.34 & 3.57 & -0.32 & 3.44\\
evi & $\beta_{18}$ & 2.07 & 1.00 & 3.28 & 1.03 & 3.11\\
sdSpatial & $\sigma$ & 1.48 & 1.01 & 2.45 & 0.91 & 2.40\\
range/1000 & $\phi_X/1000$ & 1212.73 & 450.28 & 22100.81 & 220.56 & 6667.96\\
combinedRange/1000 & $\sqrt{\phi_X \phi_Y}/1000$ & 1138.60 & 365.27 & 19530.13 & NA & NA\\
anisoRatio & $\phi_R$ & 1.13 & 1.00 & 2.08 & 0.64 & 2.00\\
shape & $\kappa$ & 0.20 & 0.07 & 0.75 & -0.14 & 0.54\\
nugget & $\nu^2$ & 0.37 & 0.00 & 1.17 & NA & NA\\
sdNugget & $\tau$ & 0.90 & 0.00 & 1.60 & 0.38 & 2.13\\
anisoAngleRadians & $\phi_{A}$ & -0.82 & -1.57 & 1.57 & -2.55 & 0.92\\
aniso1 & $\gamma_2$ & -0.02 & -0.73 & 0.71 & NA & NA\\
aniso2 & $\gamma_3$ & -0.37 & -1.00 & 0.64 & NA & NA\\
boxcox & $\lambda$ & -0.23 & -0.29 & -0.18 & -0.29 & -0.17\\
\bottomrule
\end{tabular}
\end{table}

\end{knitrout}

\section{Discussion}
Inference for anisotropic geostatistical models is challenging due to the large number of non-linear parameters involved (two range parameters, anisotropic parameters, nugget effect), and more so when the Mat\'ern shape parameter and Box-Cox parameter are treated as unknown.  Much of the software for geostatistical analysis does not have the capability to compute profile likelihoods for the model parameters, and instead  uses plug-in information-based (Wald) CIs for regression parameters (with the correlation parameters treated fixed at their MLEs).
These CIs do not take into account how uncertainty in other parameters' affects the uncertainty of $\beta$'s. Furthermore, information-based CIs are not always possible, since the information matrix is not positive definite when for example, the parameter estimate is close to a boundary. 

Computing and showing profile likelihoods can be helpful for evaluating  spatial model parameters, and likelihood-based CIs would provide wider and better coverage than information-based CIs. Obtaining profile likelihood for any of the parameters requires a large amount of repeated evaluation of the likelihoods, this computation-heavy work is more easily tackled by parallel computing with a GPU.  GPUs are designed to process thousands of parallel threads simultaneously, and are good at the repetitive likelihood evaluation and can considerably reduce the run-time. The GPU-based technique introduced here could be applied to other transformed-Gaussian models with many variance parameters, such as hierarchical models with multiple random effects, smoothing models with latent random walks, or more complex spatial models with additional random effects.

Our work focused on using dense matrix spatial algorithms without approximations, leveraging GPU's to enable the standard Cholesky-based likelihood evaluation to be performed on the reasonably large soil mercury dataset with 800 spatial locations.
The main computational limiting factor for scaling up to larger datasets is GPU memory.   Storing 400 variance matrices in double precision for 10,000 spatial locations would require 300GB of memory, which at the time of writing is only available on specialized and costly GPU hardware.  A second constraint would be the time taken to maximize the likelihood function, which  requires multiple sequential evaluations of the likelihood on the CPU.  Thirdly, large datasets with many points close to one another (relative to the range parameter) could have variance matrices which, after rounding errors, are not positive definite.  However, any dataset for which the MLE's can be obtained using the existing dense matrix Cholesky methodology could be accommodated with this GPU profile likelihood approach, provided a sufficiently powerful GPU is used.

When datasets are larger (or time constraints more severe) than a dense Cholesky algorithm can accommodate, approximations such as the SPDE approach \cite{lindgren2011explicit}, spectral approximations \cite{taylor2011lgcp}, low rank approximations \cite{banerjee2008gaussian, cressie2008fixed}, or covariance tapering \cite{furrer2006covariance, kaufman2008covariance} are necessary.  The statistical methodology of creating representative points and evaluating likelihoods in parallel is perfectly compatible with such approximations, however implementing them would require new GPU code designed to carry out the parallel computations efficiently.  The SPDE approach has become the dominant approximation for use with Bayesian inference, and a sparse matrix equivalent to the batch matrix functions in the current \pkg{gpuLik} package would enable GPU profile likelihoods for larger datasets using the SPDE model.

Extending the methodology to latent-Gaussian models, with a non-Gaussian response distribution, would be a useful but more complex undertaking.  There is no closed-form expression for the likelihood and a Laplace approximation as used by \cite{stringer2021fast} and \cite{rue2009approximate} would be the simplest approach.  Achieving this on a GPU would require an efficient batch implementation of 'inner' optimization of the joint probability $\pi(Y, U)$, which would be a substantial undertaking.   

In summary, our method leveraging parallel likelihood evaluations on the GPU is a feasible and convenient method for understanding the uncertainty in the covariance parameters of anisotropic spatial models and propagating this uncertainty through to inference on the $\beta$ regression parameters.  This includes the often disregarded Mat\'ern shape parameter, we found that although the profile likelihood of $\kappa$ is often flat, it generally has an identifiable maximum and can be profiled out.  Further, we have shown that frequentist inference using exact form of the likelihood (without approximating the variance matrix) is feasible for moderately large spatial datasets.

\section{Appendix: Restricted maximum profile likelihoods}
One problem of the maximum likelihood estimation is that variance is underestimated. Think of the simple case when the covariance matrix of $Y$ is $\sigma^2 I$, then 
\begin{align*}
	&\hat{\sigma}^2 = \frac{1}{n}(y-X\hat{\beta})^\T (y-X\hat{\beta}),\\
	&\E[\hat{\sigma}^2]=\frac{n-p}{n} \sigma^2 < \sigma^2,
\end{align*}
where $p= \text{rank}(X)$, is the number of elements in $\beta$. The estimation bias comes from not taking into account the degree of freedom used for estimating $\beta$. Restricted maximum likelihood (REML) developed by \cite{patterson1971recovery} and \cite{cox1987parameter}   is an approach that produces unbiased estimators or less biased estimators than MLE in general. REML eliminates the influence of $X$ on $\hat{\sigma}^2$ and $\hat{\omega}$. In this method, we find a matrix $A$ of full rank such that $AX=0$. The $n \times n$ projection matrix $S= I-X(X^\T X)^{-1}X^\T$ satisfies $SX=0$, however, $S$ is degenerate as $\text{rank}(S)=n-p$, so we take just $n-p$ linearly independent rows from $S$ to make the matrix $A_{(n-p)\times n}$. Transform the observations $Y$ linearly to $Y^*=AY$, then $\E(Y^*)=AX\beta=0$. Recall in \eqref{eq:1} we obtained $Y \sim N_n (X\beta, \sigma^2 V),$ thus
\begin{equation*}
	Y^* \sim N_n(0,\sigma^2 AVA^\T).
\end{equation*}
The principle of the REML method is to estimate the variance parameter $\sigma^2$ by maximizing the restricted log-likelihood function given below, where $y^*$ represents a realization of the random variable $Y^*$,
\begin{equation} \label{eq:6}
	-2\ell^*(\sigma^2, \omega ;y^*)=y^*\T(\sigma^2 AVA^\T)^{-1}y^*+\log|\sigma^2 AVA^\T|+(n-p)\log(2\pi),
\end{equation}
Differentiating \eqref{eq:6} with regard to $\sigma^2$ and setting it equal to zero gives the unbiased REML estimate for $\sigma^2$
\begin{gather*} 
	\hat{\sigma}_{reml}^2(\omega)= \frac{y^*\T (AVA^\T)^{-1} y^*}{n-p}.
\end{gather*}
The REML criterion is based on the likelihood function of $y^*$ \eqref{eq:6}, in which $X$ does not appear.  REML bypasses estimating $\beta$ and can therefore, produce unbiased estimates for $\sigma^2$.

We can write \eqref{eq:6} in terms of the observed data $y$ using the two identities derived by \cite{searle1978notebook} in Section 5.1:
$$|A (\sigma^2V)A^\T|=|\sigma^2V||X^\T (\sigma^2V)^{-1}X|$$ and $$y^*(\sigma^2 AVA^\T)^{-1}y^*=(y-X\hat{\beta})^\T (\sigma^2 V)^{-1}(y-X\hat{\beta}),$$ where $\hat{\beta}=(X^\T V^{-1}X)^{-1} X^\T V^{-1}y$.  Then the restricted log-likelihood of $y$ is
\begin{align*}
	-2\ell^*(\sigma^2, \omega;y)&=(y-X\hat{\beta})^\T(\sigma^2V)^{-1}(y-X\hat{\beta})+(n-p)\log \sigma^2+\log |V|  \notag \\  &+\log|X^\T V^{-1}X|+n\log(2\pi).
\end{align*}
If Box-Cox transformation is applied on $y$, then the general restricted log-likelihood of $y$ incorporating the Box-Cox transformation parameter would be
\begin{align}
	-2\ell^*(\sigma^2, \omega, \lambda;y)&=(y'-X\hat{\beta})^\T(\sigma^2V)^{-1}(y'-X\hat{\beta})+(n-p)\log \sigma^2+\log |V|  \notag \\  &+   \log|X^\T V^{-1}X|  +n\log(2\pi)-
	2(\lambda-1)\sum_{i=1}^{n}\log y_i,\label{eq:7}
\end{align}
which does not depend on the choice of matrix $A$. Differentiate \eqref{eq:7} with respect to $\sigma^2$ and set it equal to zero, we have the MLE $\hat{\sigma}_{reml}^2$ in terms of the transformed data $y'$ given parameters $\omega$ and $\lambda$,
\begin{equation}\label{sigmahat_reml_y}
	\hat{\sigma}_{reml}^2(\omega, \lambda)=\frac{(y'-X\hat{\beta})^\T  V^{-1}(y'-X\hat{\beta})}{n-p}.
\end{equation}
Substitute \eqref{sigmahat_reml_y} into \eqref{eq:7}, leading to the profile restricted log-likelihood for $(\omega, \lambda)$
\begin{align}\label{remlpro}
	-2\ell^*_p(\omega, \lambda;y)&=(n-p)\log \frac{(y'-X\hat{\beta})^\T  V^{-1}(y'-X\hat{\beta})}{n-p}+\log |V|+\log |X^\T V^{-1}X|  \nonumber  \\
	&-2(\lambda-1)\sum_{i=1}^{n}\log y_i +n\log(2\pi)+n-p.
\end{align}

Compared to the ``ml'' PLL given in \eqref{profile}, the effective dimension is reduced from $n$ of $\ell(\cdot)$  to $(n-p)$ of $\ell^*_p(\cdot)$ , this distinction is important if $p$ is large.

\subsection{Profile likelihood for \texorpdfstring{$\beta_p$}{Lg} using REML}
If REML method is used, then $\ell_p(\beta_i, \hat{\tilde{\beta}}_{\beta_i}, \hat{\sigma}_{\beta_i}, \omega, \lambda;y)$ would be equivalent to
\begin{align}\label{remlprofileytilde}
	-2\ell^*_p(\omega, \lambda;y)&=(n-p)\log \frac{(\tilde{y} - \tilde{X}\hat{\tilde{\beta}}_{\omega,\lambda})^\T V^{-1}(\tilde{y}-\tilde{X}\hat{\tilde{\beta}}_{\omega,\lambda})}{n-p} +\log |V|+\log |X^\T V^{-1}X| \nonumber \\ 
	&-2(\lambda-1)\sum_{i=1}^{n}\log y_i  +n\log(2\pi)+n-p.
\end{align}


\bibliographystyle{unsrt}  
\bibliography{paper2}

\end{document}